# A Greedy Link Scheduler for Wireless Networks with Gaussian Multiple Access and Broadcast Channels

Arun Sridharan, *Student Member, IEEE,* C. Emre Koksal, *Member, IEEE,* and Elif Uysal-Biyikoglu, *Member, IEEE*

## Abstract

Information theoretic Broadcast Channels (BC) and Multiple Access Channels (MAC) enable a single node to transmit data simultaneously to multiple nodes, and multiple nodes to transmit data simultaneously to a single node respectively. In this paper, we address the problem of link scheduling in multi-hop wireless networks containing nodes with BC and MAC capabilities. We first propose an interference model that extends *protocol interference model*s, originally designed for point to point channels, to include the possibility of BC and MAC. Due to the high complexity of optimal link schedulers, we introduce the Multiuser Greedy Maximum Weight algorithm for link scheduling in multi-hop wireless networks containing BCs and MACs. Given a network graph, we develop new *local pooling conditions* and show that the performance of our algorithm can be fully characterized using the associated parameter, the *multiuser local pooling factor*. We provide examples of some network graphs, on which we apply local pooling conditions and derive the multiuser local pooling factor. We prove optimality of our algorithm in tree networks and show that the exploitation of BCs and MACs improve the throughput performance considerably in multi-hop wireless networks.

## I. INTRODUCTION

The link scheduling problem for multi-hop wireless networks has received significant attention in the past few years [1]-[10]. The common assumption in these studies is point-to-point communication, that is, the possibilities of network information theory have not been incorporated. In this paper[1] we expand the scope of link schedulers to include multi-user communication scenarios using techniques developed in multi-user information theory. We first propose a generalized interference model to allow for such multi-user communication scenarios. We then introduce the Multiuser Greedy Maximum Weight (MGMW) scheduler for the proposed interference model and analyze its performance for arbitrary network graphs. For that purpose, we derive special conditions, that we shall call multiuser local pooling conditions.

In a wireless network with shared spectrum, in general, interference prevents all point-to-point nodes from being used at full capacity at the same time. The general objective of the scheduling problem is to determine which links

---

to activate simultaneously in a network. A scheduling policy is said to be throughput optimal if it can keep all the queues stable under any stabilizable arrival rate vector, that is, any arrival rate vector for which the network can be stabilized. In a multi-hop network with multiple flows and fixed link capacities, the throughput optimal scheduling problem was initially proposed in [19]. The complexity of the optimal scheme, however, is very high, making it highly impractical to implement. Recently, researchers have focused on certain classes of network interference models which impose constraints on the set of links that can be simultaneously active in a network. One such model is the *node exclusive interference model* under which a node cannot simultaneously transmit and receive, and also cannot communicate simultaneously with more than one node. An optimal schedule for the node exclusive interference model, known as Maximum Weighted Matching, has a complexity, $O(N^3)$ [16], where $N$ is the number of nodes in the network. A more general interference model is the $k-$ hop interference model, with $k$ being the the minimum number of hops between any two active links (when $k = 1$, we end up with node exclusive interference model). Maximum weight matching is NP-hard for $k \geq 2$ [9].

To address the complexity issue, low complexity suboptimal algorithms like greedy maximal scheduling have been proposed. An example of greedy scheduling is Greedy Maximal Matching (GMM) for node exclusive interference models [1]. Apart from being suitable for distributed implementation [12], GMM has the property that at each time slot the sum of the weight of the scheduled links is no less than a fraction $1/2$ of the maximum weight [8], [13]. This also leads to the conclusion that it achieves at least a fraction $1/2$ of the capacity region of the network [1]. However, the performance of the GMM scheme turns out to be far better than this lower bound in many scenarios, as shown in [14] and [6]. The authors in [6] characterized the performance of the GMM scheme using a parameter called the local pooling factor, which is obtained from the knowledge of the network topology, and interference constraints. It was shown using this local pooling factor that GMM was in fact throughput optimal for many classes of network graphs including all tree networks, under the node exclusive interference model [3].

The past work on scheduling mainly focused on orthogonal resource sharing, *i.e.,* if a link is active no other interfering link can be active simultaneously. Link models arising from the development of network or multi-user information theory have not been incorporated. For example, using *superposition coding*, a node could simultaneously transmit to two or more links at a rate lower than the individual link capacities, but higher than what could be achieved by time sharing between the individual links [15], [17]. Similarly, by using *successive interference cancellation* techniques at the receiver node in a Multiple Access Channel, two or more nodes could transmit simultaneously to a receiver node with the achievable rate region being larger than the time sharing region. In a network, nodes may form information-theoretic broadcast and multiple access channels using the appropriate multiuser encoding technique to exploit the entire capacity region of the associated multiuser channels. In this work, we design a multiuser greedy scheduling algorithm, MGMW, for networks with multiuser channels. In order to analyze the performance of MGMW, we develop an interference model and certain associated conditions, which we refer to as $\sigma_M$-local pooling conditions. These *new* conditions involve the rates achievable over the multiuser channels and are different from the classical local pooling conditions developed for the point-to-point paradigm. Based on the multiuser local pooling conditions, we derive a multiuser local pooling factor, $\sigma_M^L$, and show that the performance of our algorithm can be characterized using $\sigma_M^L$. Indeed, we show that the efficiency ratio of MGMW, defined as the largest fraction of the network stability region that can be stabilized by MGMW cannot be less than $\sigma_M^L$. We also show that the efficiency ratio of MGMW can not be larger than $\sigma_M^U$, another parameter derived from the $\sigma_M$-local pooling conditions. To illustrate our ideas clearly, we focus on networks with 2-user multiuser channels in this paper. While the generalization to the $n$-user case is straightforward, the corresponding local pooling conditions are mathematically cumbersome; consequently we do not present them in this paper.

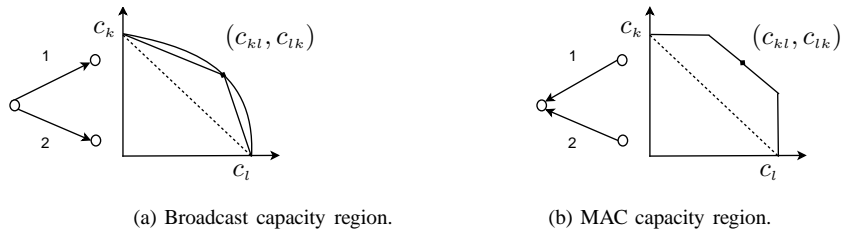

(a) Broadcast capacity region.

(b) MAC capacity region.

Figure 1: The capacity regions of the two-user Gaussian BC and MAC are illustrated in this figure.

We start the development of MGMW and its performance in Sections III and IV respectively, for multiuser channels in which the rate is a given fixed point on the multiuser capacity region. We then generalize MGMW and analyze its performance in Sections V and VI for the case in which the rate of multiuser channels is variable, chosen appropriately over the entire capacity region. A surprising conclusion we deduced from our results is that, for certain network configurations, variable-rate MGMW can lead to an inferior performance compared to fixed-rate MGMW. For both fixed-rate and variable rate MGMW, we consider examples with different network topologies to understand what topologies lead to throughput optimality for MGMW. If MGMW is not throughput optimal, we compare its performance with the optimal performance as well as the performance of GMM, the greedy scheduler without multiuser channels. We also simulate MGMW using an arbitrary network topology in Section VII.

## II. SYSTEM MODEL

We model the wireless network as a graph $\mathcal{G}(\mathcal{V}, \mathcal{E})$ where $\mathcal{V}$ is the vertex set representing the nodes and $\mathcal{E}$ is the set of edges. Each edge represents a directed point-to-point link over which a sender node transmits to a receiver node. To present ideas clearly, we assume a single-hop network traffic model. The extension to multi-hop traffic models is readily possible using techniques in [6], [4]. We assume a time slotted model indexed by $t$ in which the packets arrive at the start of every time slot. The packet arrivals in each link are independent and identically distributed across time slots, but may be correlated across links. Each node keeps a separate queue for every edge it transmits over. Let $Q_l(t)$ represent the queue length for the packets to be transmitted over edge $l$ and $\lambda_l$ denote the arrival rate for edge $l$.

We assume that a node can communicate to a single node or a *pair of nodes* simultaneously, using information theoretic broadcast and multiple access channels. The information theoretic Broadcast Channel (BC) occurs when a transmitting node sends jointly encoded data using a suitable codebook, intended to more than one receiver at the same time, from which the receivers can decode their respective information. Similarly, a Multiple-Access Channel (MAC) occurs when a node receives simultaneously from more than one transmitter nodes . The MAC (or BC) capacity region for a given power constraint is the closure of the convex hull of the set of rate vectors such that there exists codebooks at this rate, with an average power below the power constraint, such that the receivers can decode with arbitrarily small probability of error [17]. Fig. 1 illustrates the capacity region of a two user AWGN Broadcast Channel and an AWGN MAC. Given the channels, and given certain coding schemes, we have certain BC and MAC rate regions that our rates belong to. In the rest of the development, for the ease of exposition, we shall refer to these as *capacity region*s. The important observation is that the MAC and BC rate regions are strictly larger than the respective time-sharing achievable rate regions. Any point in the interior of the capacity region is achievable by choosing the appropriate set of codes. In this paper, we restrict attention to BCs and MACs comprised of only two edges, *i.e.*, a sender can transmit to two receivers (BC) or two senders can simultaneously transmit to one receiver (MAC). In the sequel, we use the term link in a generic sense to include point-to-point links as well as links made up of BC and MAC links. A *point-to-point link* consists of a sender node transmitting to a receiver node over one edge. A *multiuser link* is formed when a node transmits to two receiver nodes over a pair of edges (BC), or when a pair of nodes transmit to a common receiver node over two edges (MAC). We will specify the nature of the link wherever necessary. Note that the generalization of our results to the scenario in which a multiuser channel can involve more than two edges is straightforward. However, the algebra involved and the conditions we derive become highly complicated and we do not present this generalization in this paper.

We assume that a multiuser link $(k, l)$ can operate at any rate on the boundary or the interior of the capacity region, denoted by $\mathcal{R}_{kl}$, and that the scheduling policy has the freedom to choose the appropriate rate point $(c_{kl}(t), c_{lk}(t))$ in each time slot. In Section III, we initially assume a simpler network model where each transmitter, capable of multiuser communication chooses a single rate pair from the boundary of the capacity region for the associated multiuser link (BC or MAC), and whenever it chooses to transmit over that link, it uses the associated rate pair. Thus, for a multiuser link $(k, l)$, $(c_{kl}(t), c_{lk}(t)) = (c_{kl}, c_{lk})$ is fixed throughout and the scheduler does not have the freedom to choose the rate of the multiuser link. In Fig. 1, $(c_{kl}, c_{lk})$ is the fixed rate pair associated with the multiuser link $(k, l)$. In Section V, we relax this assumption so that the multiuser link can operate at any point in the capacity region, with the scheduling policy having the freedom to choose an appropriate rate point every time slot. Note that the point-to-point link capacities also belong to the multiuser link capacity region (as $(c_k, 0)$ and $(0, c_l)$), but are defined separately to distinguish these from the multiuser link case, where both edges of the multiuser link have non-zero transmission rates. For a Gaussian BC or MAC, as seen in Fig. 1, $c_{kl} < c_k$, and $c_{lk} < c_l$ and $\frac{c_{kl}}{c_k} + \frac{c_{lk}}{c_l} > 1$. This means that we can achieve rates strictly larger than those achieved by time sharing between the two point-to-point links. It is worth stressing that using a BC link does not mean that the same message is being sent to both receivers, but rather different messages (packets) are sent to the receivers connected by the edges $l$ and $k$. Similarly, using a multiple-access channel entails the two transmitters sending different information to the common receiver. It should also be noted that we do not rule out the possibility of any edge being utilized as a point-to-point link, even if it is a part of some multiuser link.

To incorporate the possibility of such information theoretic BC and MAC links, we are motivated to introduce a generalized binary interference model. Similar to classical binary interference models, each edge $l \in \mathcal{E}$ is associated with a set consisting of all edges that conflict with $l$, *i.e.,* the links that absolutely cannot be scheduled when edge $l$ is scheduled. We call this set the *main interference set* and denote it by $X_l$. For edge $l$, let $Y_l$ denote the set of edges that can be paired with $l$ to form a multiuser link. We call $Y_l$ the *secondary interference set* of $l$, to distinguish it from the main interference set, and also due to the fact that two edges that lie in each other's secondary interference set do not necessarily exclude each other: they simply reduce each other's rates. Note that $Y_l$ and $X_l$ are mutually exclusive sets and if edge $k \in Y_l$, then $l \in Y_k$. Let $c_l$ be the individual capacity of the point-to-point link $l$. Link $l$ can be active at rate $c_l$ only if no other edge $k \in X_l \cup Y_l$ is active. If edge $k \in Y_l$ is active simultaneously as edge $l$, then it implies that they are active as multiuser link $(k, l)$, at some rate $(c_{kl}(t), c_{lk}(t))$, chosen from the boundary of the capacity region, as illustrated in Fig. 1. We also observe that the notion of *main* and *secondary* interference sets could serve a more general purpose than allowing for multiuser links. For instance, it is possible to extend the definition of secondary interference sets to include Interference Channels [20], a scenario in which two or more interfering links can be active simultaneously at reduced rates using suitable coding techniques. We use the term interfering links to indicate that these links are not allowed to be simultaneously active in a protocol interference model. Our interference model, while incorporating multiuser links, does carry over some limitations of the protocol interference model, namely the discrete nature of interference. The rate of a link ideally depends on the interference caused by other links, and this is captured in more realistic models like the SINR model. Scheduling problem with the SINR model has also been investigated extensively in the literature [10]-[15]. However, the problem of optimal scheduling with the SINR model is known to be NP hard and this fact is one of the main motivations of the use of approximate discrete interference models such as the binary interference model. Owing to a significant portion of the literature focusing on such models, there is a better understanding of wireless scheduling for graph based or binary interference models. We leverage this understanding in our extension of the protocol interference model when we introduce the idea of secondary interference sets. Further, the capacity regions of the multiuser links in the presence of interfering links may also not be known. However, our model only requires knowing an achievable rate region for the multiuser link which is strictly convex.

We define a rate allocation vector $\vec{r}^{1 \times |\mathcal{E}|}$ of link rates where $\vec{r}_l$ represents the rate of transmission over the edge $l$. A rate allocation vector must satisfy the following constraints:

(i) If $r_l > 0$ then $r_k = 0, \ \forall k \in X_l$. This condition describes the main interference constraint for a point-to-point link $l$.

(ii) If $r_l > 0$ and $r_k > 0$, and also if $k \in Y_l$ and $l \in Y_k$, then $r_l = c_{lk}$, and $r_k = c_{kl}$. Furthermore, $r_j = 0$ for all $j \in Y_l \cup Y_k$ where $j \neq k, l$ . This condition captures the constraints arising from the secondary interference set $Y_l$: If a multiuser link is scheduled then the edges belonging to the secondary interference sets of either of the two edges that constitute the multiuser link cannot be scheduled. Thus a node is allowed to transmit or receive simultaneously over at most two edges.

(iii) There exists no $j \in \mathcal{E}$ such that $j$ does not interfere with any link and yet is not scheduled.

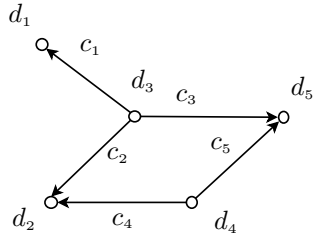

Figure 2: Five edge network with point-to-point link capacities $c_1, c_2, c_3, c_4$ and $c_5$. Links (1,2) and (4,5) each form broadcast links.

Let R denote the set of all possible rate allocation vectors $\vec{r}$ on a network graph. In general, R can be an uncountable set, since it includes rate allocation vectors corresponding to every rate pair on the boundary of the capacity region of a multiuser link. For any subset $E \subset \mathcal{E}$, $\mathsf{R}_E$ is defined as the set of all $|E|$-dimensional rate allocation vectors that satisfy the constraints in (i), (ii) and (iii), with $E$ substituted for $\mathcal{E}$ in constraint (iii). As an example, a network graph comprised of five edges is shown in Fig. 2, along with the set of feasible rate vectors. Table I describes the interference sets that we could define for this network under a node-exclusive interference model. In the figure, node $d_3$ can set up a multiuser link to send data to $(d_1, d_2)$ at a rate $(c_{12}, c_{21})$, chosen from the capacity region. Similarly, node $d_4$ can transmit to $(d_2, d_5)$. The rate allocation vectors $\vec{r}$ for this network are: $[0\ c_2\ 0\ 0\ c_5]$, $[0\ 0\ c_3\ c_4\ 0]$, $[c_1\ 0\ 0\ 0\ c_5]$, $[c_1\ 0\ 0\ c_4\ 0]$, $[c_{12}\ c_{21}\ 0\ 0\ c_5]$, and $[c_1\ 0\ 0\ c_{45}\ c_{54}]$, where $(c_{12}, c_{21})$ and $(c_{45}, c_{54})$ are non zero rate pairs chosen from their respective multiuser capacity regions denoted by $\mathcal{R}_{12}$ and $\mathcal{R}_{45}$. Note that in the absence of the BC links only the first four rate vectors would be available.

The *optimal stability region, or equivalently, the capacity region of a network* is the set of all arrival rate vectors such that for any arrival vector in this set, there exists some scheduling scheme that can keep the queue lengths from growing unbounded. Here, we use the term optimal stability region to distinguish the network capacity region from the multiuser information theoretic capacity region. The optimal stability region of the network [19] is given by the interior of the set $\Lambda = \{\vec{\lambda} : \vec{\lambda} \preceq \vec{\phi}, \text{ for some } \vec{\phi} \in Co(\mathsf{R})\}$, where $Co(\mathsf{R})$ denotes the convex hull of the vectors in R and $\preceq$ represents componentwise inequality. Let $\pi : \vec{Q}(t) \to \mathsf{R}$ be a scheduling policy that selects a feasible rate vector for every time slot, based on the queue length state vector $\vec{Q}(t)$. Let $\Pi$ denote the set of all such scheduling schemes $\pi$. For this model, the entire capacity region can be achieved by the Maximum Weight scheduler [19], which at every time slot $t$, selects the rate vector which has the highest sum of queue-weighted rates. To compare the advantage of using multiuser links, we also define a set of scheduling policies that cannot utilize the possibility of multiuser links. Let $\hat{\mathsf{R}}$ denote the set of rate vectors for a network with sole point-to-point communication. Set $\hat{\mathsf{R}}$ satisfies the following interference constraints: $r_i > 0 \Rightarrow r_k = 0, \ \forall k \in X_i \cup Y_i$. Let $\hat{\pi} : Q(t) \to \hat{\mathsf{R}}$ be a scheduling scheme designed based on this constraint and let $\hat{\Pi}$ denote the set of all such schemes $\hat{\pi}$. Note that $\hat{\Pi} \subset \Pi$. Our objective is to find a low complexity scheme that belongs to the set $\Pi$ and characterize its performance with respect to the capacity region, as well as to compare its performance to that of other schemes chosen solely from $\hat{\Pi}$. In the next section, we describe the MGMW scheme for the network model in which the multiuser link rates are a fixed point on the boundary of the capacity region.

| link $l$ | $X_l$ | $Y_l$ |
|---|---|---|
| 1 | {3} | {2} |
| 2 | {3,4} | {1} |
| 3 | {1,2,5} | $\emptyset$ |
| 4 | {2} | {5} |
| 5 | {3} | {4} |

Table I: Interference sets for the five edge network of Fig. 2.

## III. MULTIUSER GREEDY MAXIMUM WEIGHT (MGMW) ALGORITHM

Let the rate of any multiuser link $(k, l)$ be a fixed point $(c_{kl}, c_{lk})$ on the boundary of the capacity region so that $c_{kl}(t) = c_{kl}$ and $c_{lk}(t) = c_{lk}$. In this case, the set of all rate allocation vectors R is now a finite set. For the network model with fixed multiuser link rates, we present a "greedy" scheduling policy, MGMW which selects a rate allocation vector from R in each time slot. MGMW, in principle is similar to the GMM, discussed in [6]. Before giving a precise definition of MGMW, it will be instructive to summarize its operation descriptively:

Each link is assigned a weight, which is basically the queue-weighted link rates. In each time slot, MGMW first greedily picks the link (point-to-point or multiuser) with the highest weight. It then removes all interfering links and picks the link with the highest weight from the remaining links. This process goes on until there are no more links left to pick. More precisely, let $\mathcal{L}$ denote the set of all links (point-to-point as well as multiuser), *i.e.,*

$$\mathcal{L} = \{\mathcal{E} \cup \{(k, l) \in \mathcal{E}^2 \mid k \in Y_l \text{ and } l \in Y_k\}\}.$$

For any element $m \in \mathcal{L}$, we define the weight of a link $W_m(t)$ as follows:

$$W_m(t) = \begin{cases} Q_j(t)c_j, & m \text{ is a point-to-point link } j \\ Q_k(t)c_{kl} + Q_l(t)c_{lk}, & m \text{ is a MAC/BC link } (k, l) \end{cases}. \tag{1}$$

MGMW operates as follows. At any point in the algorithm, let $Z$ denote the set of currently unselected links that do not interfere with any of the selected links. MGMW initializes $Z$ to $\mathcal{L}$ and repeats steps $1 - 2$ until $Z = \emptyset$.

1. Select a link $m$ with the highest weight in $Z$.

$$m \in \underset{n \in Z}{\mathrm{argmax}}\{W_n(t)\}. \tag{2}$$

   Note that $m$ need not be unique. In case of a tie between a point-to-point link and a multiuser link, MGMW gives priority to the point-to-point link.

2. After the selection, remove all links that conflict with $m$, *i.e.,* set their rates in the rate allocation vector to zero. If $m$ is a point-to-point link $j$ then the scheduler sets $r(k) = 0$, for all $k \in X_j \cup Y_j$. If $m$ is a multiuser link $(k, l)$ then it sets $r(i) = 0$, for all $i \in X_k \cup Y_k \cup X_l \cup Y_l$ except $i = k, l$. Update $Z$ to consist of only non-interfering links.

At the end of the procedure MGMW yields a rate vector that belongs to the set R. Also, if $Y_l = \emptyset, \forall l \in \mathcal{E}$, MGMW reduces to the GMM of [6].

**Example 1.** . MGMW for the five edge network of Fig. 2.

The set of point-to-point, and multiuser links for this case is given by $\mathcal{L} = \{1, 2, 3, 4, 5, (1, 2), (4, 5)\}$. Let the link rates be $c_1 = 4$, $c_2 = 6$, $c_3 = 2$, $c_4 = 8$, $c_5 = 5$, $(c_{45}, c_{54}) = (4, 3)$ and $(c_{12}, c_{21}) = (3, 4)$. Let $Q_1(t) = 20$, $Q_2(t) = 5$, $Q_3(t) = 2$, $Q_4(t) = 12$ and $Q_5(t) = 1$. Applying the MGMW algorithm, link 4 is observed to have the highest weight of 96. Link 1 as well as $(1, 2)$ each have weight 80. Link 4, having the highest weight, is picked first and following step 2, the interfering edges 2 and 5 given in Table I are removed from set $Z$. Among the remaining links, highest weight is seen to be 80 for link 1. Node $d_3$ is hence selected to transmit over link 1. The chosen rate allocation vector is then $[4\ 0\ 0\ 8\ 0]$. Thus, at time $t$, no multiuser link is chosen to transmit.

### A. Performance Characterization of MGMW Scheduler

We adopt the definition of efficiency ratio given in [6] to describe the performance of the MGMW algorithm. The efficiency ratio of the MGMW scheduling algorithm $\gamma$ is defined as the largest fraction of the capacity region such that any arrival rate vector inside this region can be stabilized by MGMW, *i.e.*,

$$\gamma^* := \sup \{\gamma \mid \text{the system is stable under MGMW} \tag{3}$$
$$\text{for all arrival rate vectors } \vec{\lambda} \preceq \gamma\Lambda.\}$$

We study the efficiency of the MGMW algorithm for any network by relating it to a parameter that we call the *multiuser local pooling factor*, which depends on the network topology and the interference sets. In the no multiuser link scenario, i.e, when the set $Y_l = \emptyset$ for all $l$, [14] showed that the GMM scheduler is throughput optimal for network graphs which satisfy certain conditions. These conditions, known as local pooling conditions are based on the network topology and the link interference constraints. In [6], a more general condition called $\sigma$-local pooling was introduced to characterize the performance of GMM for arbitrary interference graphs, including those for which GMM was not throughput optimal. s In this section we identify new network conditions in the presence of multiuser links, which we call *multiuser local pooling ($\sigma_M$-local pooling) conditions*. We will use these conditions to define the multiuser local pooling factor for any network graph. Recall that $\mathcal{L}$ is the set of all links for a given network graph. To describe the $\sigma_M$-local pooling conditions, we focus on certain subsets of $\mathcal{L}$, which we call candidate maximum weight (MW) subsets. A set of links $L_{MW} \subset \mathcal{E}_B$ is called a candidate MW subset, if there exist queue lengths, not all zero, such that $L_{MW} = \operatorname{argmax}_{n \in Z}\{W_n\}$. Not all subsets of $\mathcal{E}_B$ can be candidate MW subsets. In fact, every candidate MW subset $L_{MW}$ satisfies the following property:[2] For any pair of edges $k$ and $l$ such that $k \in Y_i$, both individual edges $k$ and $l$ of the multiuser link $(k, l)$ do not appear as point-to-point links in $L_{MW}$ separately. In other words, if an edge $j$ appears as a point-to-point link in the set $L_{MW}$, then no other edge in the secondary interference set of $j$ appears as a point-to-point link in $L_{MW}$: $\{j \in L_{MW} \Rightarrow l \notin L_{MW} \text{ for all } l \in Y_j\}$. We also denote $\mathcal{E}_{L_{MW}}$ to be the set consisting of all the edges in $L_{MW}$.

---

[2]Consider any $L_{MW}$. Suppose there are two point-to-point links $k \in L_{MW}$, $l \in L_{MW}$ such that $l \in Y_k$. Then the weights of the links $k$ and $l$ are equal, i.e, $q_k c_k = q_l c_l$. Suppose the weight of the broadcast link is not greater than that of the individual links, $q_k c_{kl} + q_l c_{lk} \leq q_k c_k$. Substituting for $q_l$ from $q_k c_k = q_l c_l$ gives us the condition $\frac{c_{kl}}{c_k} + \frac{c_{lk}}{c_l} \leq 1$. This contradicts our earlier assumption on the rates for the broadcast channel that $\frac{c_{kl}}{c_k} + \frac{c_{lk}}{c_l} > 1$, and hence the weight of the broadcast link exceeds that of the individual links. Thus, $L_{MW}$ cannot be the set with the highest weight.

**Example 2.** Consider the network graph of Fig. 2. Examples of candidate MW subsets for this graph are the sets $\{1, 3, (4, 5)\}$ and $\{(1, 2), 2, 3, (4, 5)\}$. The sets $\{(1, 2), 1, 2, 3\}$ and $\{1, 2, 3\}$ however, are not candidate MW subsets, as the links 1 and 2 appear together as point-to-point links in both sets, while also comprising the edges of the multiuser link $(1, 2)$. For this network graph the set $\mathcal{E}_{L_{MW}} = \{1, 2, 3, 4, 5\}$.

We now introduce the idea of $\sigma_M$-local pooling conditions applied to candidate MW subsets.

**Definition 3.** Let $L_{MW}$ be any candidate MW subset. Then $L_{MW}$ contains point-to-point and multiuser links as its elements. Let $\mathsf{R}_{L_{MW}}$ denote the set of all rate allocation vectors for the set $\mathcal{E}_{L_{MW}}$. Also, let $\tilde{\mathsf{R}}_{L_{MW}} \subseteq \mathsf{R}_{L_{MW}}$ be the set of rate allocation vectors on $\mathcal{E}_{L_{MW}}$, that can be chosen by the MGMW policy, when links in $L_{MW}$ have the maximum weight. Then $L_{MW}$ satisfies $\sigma_M$- local pooling if, for any given pair $\vec{\mu}, \vec{\nu}$, where $\vec{\mu}$ is a convex combination of the rate vectors in $\mathsf{R}_{L_{MW}}$ and $\vec{\nu}$ is a convex combination of the rate vectors in $\tilde{\mathsf{R}}_{L_{MW}}$, either of the following hold:

(i) There exists a point-to-point link $j \in L_{MW}$ such that $\sigma_M \mu_j < \nu_j$, or

(ii) There exists a multiuser link $(k, l) \in L_{MW}$ such that $\sigma_M (\mu_k c_{kl} + \mu_l c_{lk}) < \nu_k c_{kl} + \nu_l c_{lk}$.

Condition (i) becomes the standard $\sigma$-local pooling condition of [6], when defined for an arbitrary subset of edges in $\mathcal{E}$. The $\sigma_M$-local pooling condition is distinguished by the fact that it is stated only over candidate MW subsets. We introduce (ii) to generalize it to the case where multiple-edge links are possible, such as information theoretic broadcast or multiple access channels. We define a parameter, $\sigma_M^L$, for a network as the supremum of all $\sigma_M$ such that every candidate MW subset $L_{MW}$ of $\mathcal{L}$ satisfies $\sigma_M$-local pooling, *i.e.*,

$$\sigma_M^L = \sup \{\sigma_M \mid \forall L_{MW} \in \mathcal{L}, \text{ conditions (i) or (ii)are satisfied for every} \vec{\mu} \text{ and } \vec{\nu}\},$$

where $\vec{\mu}$ and $\vec{\nu}$ are convex combinations of the rate vectors in $\mathsf{R}_{L_{MW}}$ and $\tilde{\mathsf{R}}_{L_{MW}}$ respectively. We call $\sigma_M^L$ a multiuser local pooling factor. To show throughput optimality when local pooling conditions were satisfied, the authors in [14] argued that if a set of links alternately had the highest queue weighted rate in a small interval of time, and if they satisfied local pooling, then GMM served to bring down the highest weights in that interval. The proof used a fluid limit argument to find a Lyapunov function whose drift was then shown to be negative. A similar approach is followed in the proof of Lemma 1 in [6]. When multiuser links are included, one needs to consider the weight of both point-to-point and multiuser links. This leads to local pooling conditions being defined over a fixed class of subsets, i.e., candidate MW subsets of links and not over all subsets of links. The reason for this will become evident in the proof of Theorem 1, where it is seen that while considering the set with links having maximum weight, one may exclude non-candidate MW subsets, as links in these sets cannot have the maximum weight simultaneously.

Based on the $\sigma_M$ local pooling conditions, we identify a condition for a candidate MW subset $L_{MW}$ with which $\sigma_M$-*local pooling* does not hold. $L_{MW}$ does not satisfy $\sigma_M$-*local pooling* if there exists a pair of vectors $\vec{\mu}, \vec{\nu}$ that are convex combinations of the rate vectors in $\mathsf{R}_{L_{MW}}$ and $\tilde{\mathsf{R}}_{L_{MW}}$ respectively, such that they satisfy $\sigma_M \vec{\mu} \succeq \vec{\nu}$. We define the parameter $\sigma_M^U$ as follows:

$$\sigma_M^U = \inf\{\sigma_M \mid \exists \text{ an } L_{MW} \in \mathcal{L}, \text{ such that } \sigma_M \vec{\mu} \succeq \vec{\nu} \text{ for some } \vec{\mu}, \vec{\nu}\}.$$

where $\vec{\nu}$ is a convex combination of rate vectors in $\tilde{\mathsf{R}}_{L_{MW}}$ and $\vec{\mu}$ is a convex combination of rate vectors in $\mathsf{R}_{L_{MW}}$. We now give Theorem 1 and Theorem 2 to prove that the efficiency ratio of the MGMW scheduler satisfies the relation $\sigma_M^L \leq \gamma^* \leq \sigma_M^U$.

**Theorem 1.** *The network is stable under MGMW algorithm for all arrival rate vectors $\vec{\lambda}$ satisfying $\vec{\lambda} \in \sigma_M^L \Lambda$.*

   *Proof:* The Proof is given in Appendix A. ∎

While Theorem 1 shows that any arrival rate within $\sigma_M^L \Lambda$ is stabilizable by the MGMW algorithm, we further link the performance of the MGMW to $\sigma_M^U$ in Theorem 2 by showing that there exist arrival rates, arbitrarily close but strictly outside of $\sigma_M^U \Lambda$, for which the system is unstable under the MGMW scheme. Theorem 1, together with Theorem 2 implies that the efficiency ratio of MGMW is bounded below by $\sigma_M^L$ and bounded above by $\sigma_M^U$.

**Theorem 2.** *Let there exist a candidate MW subset $L_{MW} \in \mathcal{L}$ such that for some positive number $\sigma_M$, and a pair of vectors $\vec{\mu}, \vec{\nu}$, which are convex combinations of the elements in $\mathsf{R}_{L_{MW}}$ and $\tilde{\mathsf{R}}_{L_{MW}}$, $\sigma_M \vec{\mu} \succeq \vec{\nu}$ is satisfied. Then, for any $\epsilon > 0$, there exists a $\vec{k} \succ 0$ such that the arrival rate $\vec{\lambda} = \vec{\nu} + \epsilon \vec{k}$ makes the system unstable under the MGMW scheme.*

   *Proof:* The proof is given in Appendix B. ∎

Theorem 1 and Theorem 2 establish sufficient and necessary conditions respectively for local pooling to occur under the MGMW policy. Unlike the GMM policy, for which a single condition is both necessary and sufficient for local pooling, we have separate necessary and sufficient conditions due to the presence of multiuser links. While these two conditions are identical in certain cases such as tree networks satisfying specific rate constraints, they are not necessariry identical in a general network.

## IV. Performance of MGMW

In this section, we analyze the performance of MGMW in some sample network topologies. We use the bounds on efficiency ratio of MGMW obtained in the previous section to evaluate the throughput gain by leveraging multiuser links in these network graphs. Even though the optimal stability region of a network with multiuser links is larger than that of a network with point-to-point links alone, for some networks it may be possible that MGMW achieves a smaller stability region than GMM. However, we provide examples of two network graphs, specifically a tree network graph and a star network graph, for which we show that MGMW achieves a larger stability region than GMM. In this process, we also explore the tightness of our bounds. In particular, we show that the lower bound is tight for tree networks when the multiuser link rates are restricted to certain rate regions. Moreover, the lower bound on efficiency ratio is good enough to prove that the stability region with MGMW is larger than that with GMM for the star network example. We also simulate the performance of MGMW and GMM on a randomly generated network graph, and observe that MGMW can stabilize a higher arrival rate for each link in the random network graph.

We first show that MGMW is throughput optimal for certain tree networks.

### A. Tree Networks

In the following theorem, we show that under the node exclusive interference assumption, MGMW is throughput optimal for directed tree graphs in which no two multiuser links have a link in common. The node exclusive interference assumption for our model only restricts a node from transmitting and receiving at the same time. It does not restrict a node from transmitting simultaneously to multiple nodes, or receiving simultaneously from multiple nodes.

**Theorem 3.** *Let $\mathcal{G} = (\mathcal{V}, \mathcal{E})$ be a directed tree graph such that $Y_k \cap Y_l = \emptyset$ for all $k, l \in \mathcal{E}$. Let $c_{mn}^2 + c_{nm}^2 > \max(c_m c_{mn}, c_n c_{nm})$ for every multiuser link $(m, n)$ in $\mathcal{L}$. Then, if the primary and secondary interference sets are constructed under the node exclusive interference assumption, $\sigma_M^L = \sigma_M^U = 1$, implying that MGMW is throughput optimal for this tree network graph.*

The proof is given in Appendix C. Theorem 3 also shows that MGMW is throughput optimal for downlink cellular networks without intercell interference with multiuser links consisting of broadcast channels. This is because the downlink cellular model is an instance of a tree network.

### B. A Network with $\sigma_M^U < 1$.

Consider the network graph shown in Fig. 3. All links have a rate of 1 when used as point-to-point links. All multiuser links have a rate of $(0.75, 0.75)$ each. We define $\sigma_{L_{MW}}$ for a candidate MW set $L_{MW}$ as the highest value of $\sigma_M$-local pooling satisfied by $L_{MW}$. We show in [21] that $\sigma_{L_{MW}} \geq 2/3$ for all candidate MW subsets $L_{MW}$ of the star network graph in Fig. 3, and hence the efficiency ratio of MGMW is at least $2/3$ for this network graph. Here we will only show this for one candidate MW subset and will not repeat the same operation for all candidate MW subsets. In order to find $\sigma_{L_{MW}}$, we make use of the following relation that we derive in Appendix C:

$$\sigma_{L_{MW}} \geq \frac{\min_{i \in 1 \ldots |\tilde{\mathsf{R}}_{L_{MW}}|} \|\vec{\tilde{r}}^i \mathbf{T}\|_1}{\max_{j \in 1 \ldots |\mathsf{R}_{L_{MW}}|} \|\vec{r}^j \mathbf{T}\|_1}, \tag{4}$$

where $\vec{\tilde{r}}^i \in \tilde{\mathsf{R}}_{L_{MW}}$, $\vec{r}^j \in \mathsf{R}_{L_{MW}}$ and $T$ is a $|\mathcal{E}_{L_{MW}}| \times |\mathcal{E}_{L_{MW}}|$ matrix such that $T_{ii} = 1$, if $i \in L_{MW}$; $T_{lk} = c_{kl}$ and $T_{ll} = c_{lk}$ if $(k, l) \in L_{MW}$; and $T_{ij} = 0$ otherwise.

Consider the candidate MW subset $L_{MW} = \{(1, 2), (3, 4), (5, 6), (7, 8), (9, 10), (11, 12)\}$. The set $\tilde{\mathsf{R}}_{L_{MW}}$ is given

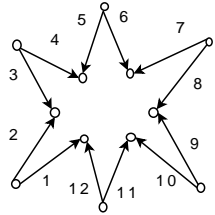

Figure 3: A twelve edge network with six broadcast links.

by:

$$
\begin{bmatrix}
\vec{\nu}_1 \\
\vec{\nu}_2 \\
\vec{\nu}_3 \\
\vec{\nu}_4 \\
\vec{\nu}_5
\end{bmatrix}
=
\begin{bmatrix}
\frac{3}{4} & \frac{3}{4} & 0 & 0 & \frac{3}{4} & \frac{3}{4} & 0 & 0 & \frac{3}{4} & \frac{3}{4} & 0 & 0 \\
0 & 0 & \frac{3}{4} & \frac{3}{4} & 0 & 0 & \frac{3}{4} & \frac{3}{4} & 0 & 0 & \frac{3}{4} & \frac{3}{4} \\
\frac{3}{4} & \frac{3}{4} & 0 & 1 & 0 & 0 & \frac{3}{4} & \frac{3}{4} & 0 & 1 & 0 & 0 \\
0 & 0 & \frac{3}{4} & \frac{3}{4} & 0 & 1 & 0 & 0 & \frac{3}{4} & \frac{3}{4} & 0 & 1 \\
0 & 1 & 0 & 0 & \frac{3}{4} & \frac{3}{4} & 0 & 1 & 0 & 0 & \frac{3}{4} & \frac{3}{4}
\end{bmatrix}
$$

Applying the relation (32) to the specified $L_{MW}$ set yields $\sigma_{L_{MW}} \geqslant 2/3$. Furthermore, we also show in [21] that $\sigma_{L_{MW}} \leq 3/4$ for this network graph. For the same network graph, if we exclude the possibility of multiuser links, it can be shown that GMM algorithm has an efficiency ratio of 2/3 (with respect to the capacity region in the no multiuser link scenario). To see this, we only need to consider the set $L = \{1, 2, 3, 4, 5, 6, 7, 8, 9, 10, 11, 12\}$. The set $L$ forms a cycle, and in the same manner as that shown for the six-link cycle network in [6], one obtains the local pooling factor as 2/3. Hence, in this example, MGMW guarantees a larger stability region compared to GMM. This is because the optimal stability region of a network for the multi-user case is a superset of the optimal stability region of the same network in the no multiuser link scenario.

In the following section, we explore a more general network model, where we assume that rather than just one point, the whole rate region is available to the encoder and decoder, as well as the scheduler. The scheduler then selects a suitable multiuser link rate pair from this region in every time slot. For this network model, we provide a generalized version of the MGMW scheme and analyze its performance by deriving local pooling conditions that are similar to the ones derived for the case with fixed multiuser link rates.

## V. Using the entire multiuser capacity region: variable rate MGMW

We present the extension of the MGMW scheme wherein we allow it to select an arbitrary rate pair from the rate region of each multiuser link in each time slot. From this point on, we restrict attention to network graphs that only have BC links as multiuser links, and omit the technical treatment of MAC links due to space constraints. While the variable rate MGMW algorithm remains unchanged in the presence of MAC links, the performance analysis of variable rate MGMW involves treating the MAC links and BC links as separate cases.

The MGMW scheme, as described before in Section III, initially determines the weight of all point-to-point and multiuser links. Since the entire multiuser capacity region is now available to the scheduler, the weight of a multiuser link is defined as the sum of queue weighted edge rates, maximized over all possible choices of rate pairs within the capacity region of the multiuser link. For any element $m \in \mathcal{L}$, We define the weight of a link $W_m(t)$ as follows:

$$W_m(t) = \begin{cases} Q_j(t)c_j, & m \text{ is a point-to-point link } j \\ \max\limits_{(c_{kl}, c_{lk}) \in \mathcal{R}_{kl}} Q_k(t)c_{kl} + Q_l(t)c_{lk}, \\ & m \text{ is a multiuser link } (k,l), \end{cases} \tag{5}$$

where $R_{kl}$ is the capacity region of the multiuser link $(k,l)$, and is assumed to be known. For instance, consider the operation of variable rate MGMW scheme for the network graph of Fig. 2. $\mathcal{R}_{12}$ and $\mathcal{R}_{45}$ are the multiuser rate regions of links $(1,2)$ and $(4,5)$ and are assumed to be known. The variable rate MGMW then uses (5) to calculate the weight of each link. Let $(c_{45}^\star(t), c_{54}^\star(t))$ and $(c_{12}^\star(t), c_{21}^\star(t))$ be the transmission rate pairs that correspond to the weight of the multiuser links $(1,2)$ and $(4,5)$. Now that all the weights and their associated transmission rates are known, variable rate MGMW then assigns the rate allocation vector following the same procedure as described for the MGMW scheme in Section III.

For a BC link, since the capacity region of the multiuser link is strictly convex, as shown in Fig. 1, the transmission rate pair associated with the weight of the link is unique, and corresponds either to a multiuser link configuration (if the rate pair that yields the maximum weight allocates non-zero rates to both edges) or to one of the point-to-point links (if the rate of one of the edges is zero). Thus, only one of the configurations can be associated with the weight of the multiuser link at any given time. With the new weights calculated as described above, the operation of the MGMW algorithm is identical to the one described in Section III, and yields a rate allocation vector that belongs to the set R. The set R for a network graph is now an uncountable set, as it includes every rate pair on the boundary of the capacity region, rather than one fixed rate point.

*Local pooling conditions for variable rate MGMW*

The performance of the variable rate MGMW scheme can be analyzed in a manner similar to that of MGMW, by relating the multiuser $\sigma_M$-local pooling factor to the efficiency ratio of variable rate MGMW. We retain the same notion of efficiency introduced earlier in Section IV. Note that the network capacity region defined in Section II is a superset of the capacity region obtained by fixing the multiuser link rates. We now present the multiuser $\sigma_M$-pooling pooling conditions for the variable rate MGMW scheme.

Let $L_{MW}$ be any candidate MW subset, *i.e.,* a subset of $\mathcal{L}$ whose links can achieve the maximum weight simultaneously. Every candidate MW subset $L_{MW}$ satisfies the following property: For any pair of edges $k$ and $l$ such that $k \in Y_l$ and vice versa, a candidate MW subset $L_{MW}$ can contain at most one out of the three links $(k,l)$, $k$ and $l$, i.e.,

$$\mathbf{I}_{(k,l) \in L_{MW}} + \mathbf{I}_{k \in L_{MW}} + \mathbf{I}_{l \in L_{MW}} \le 1,$$

for any multiuser link $(k,l)$, where $\mathbf{I}_{n \in L_{MW}}$ is an indicator variable taking the value 1 when $n \in L_{MW}$ and 0 otherwise. This property of $L_{MW}$ follows from the fact that the weight of link $(k,l)$, as a consequence of being maximized over the entire multiuser capacity region $\mathcal{R}_{kl}$, can correspond to only one out of the three possible configurations (point-to-point link $k$, point-to-point link $l$ or multiuser link $(k,l)$).

Given a candidate MW subset, the transmission rates of the point-to-point links are fixed at their respective capacity values, whereas the transmission rates of multiuser links can take different values from the associated capacity regions. Let $\mathcal{C}_{L_{MW}}$ be the set containing all possible transmission rates associated with the links in $L_{MW}$. We illustrate these sets in the example below:

**Example 4.** Consider the network graph of Fig. 2. An example of a candidate MW subset for this graph is the set $\{1, 3, (4, 5)\}$. The sets $\{(1, 2), 1, 3\}$ and $\{(1, 2), 2, 3\}$ however, are not candidate MW subsets, as they both contain links (links $(1, 2)$ and $1$ in one, and $(1, 2)$ and $2$ in the other) that cannot have the maximum weight at the same time. For the example $L_{MW} = \{1, 3, (4, 5)\}$, $\mathcal{E}_{L_{MW}}$ is $\{1, 3, 4, 5\}$ and the edges have rates given by $\{c_1, c_3, (c_{45}^*, c_{54}^*)\}$, where $(c_{45}^*, c_{54}^*)$ is an arbitrary non-zero rate pair selected from the boundary of $\mathcal{R}_{45}$. Since the rates of point-to-point links are fixed, each element of $\mathcal{C}_{L_{MW}}$ would correspond to a non-zero rate pair chosen from $\mathcal{R}_{45}$.

We now state the $\sigma_M$-local pooling conditions applied to candidate MW subsets.

**Definition 5.** *Let* $\mathsf{R}_{L_{MW}}$ *be the set of all rate allocation vectors on the subset of edges* $\mathcal{E}_{L_{MW}}$. *For each transmission rate vector* $\vec{c} \in \mathcal{C}_{L_{MW}}$, *let* $\mathsf{R}_{L_{MW}}^{\vec{c}}$ *be the set of all rate allocation vectors that MGMW can assign on* $\mathcal{E}_{L_{MW}}$, *when the links in* $L_{MW}$ *achieve maximum weight with their transmission rates given by* $\vec{c}$. *Note that* $\mathsf{R}_{L_{MW}}^{\vec{c}} \subset \mathsf{R}_{L_{MW}} \; \forall \vec{c} \in \mathcal{C}_{L_{MW}}$. *We then say that* $L_{MW}$ *satisfies* $\sigma_M$-*local pooling, if every* $\mathsf{R}_{L_{MW}}^{\vec{c}}$ *satisfies the following conditions:*
*For any two vectors* $\vec{\mu}$ *and* $\vec{\nu}$, *where* $\vec{\mu}$ *is a convex combination of elements in* $\mathsf{R}_{L_{MW}}$, *and* $\vec{\nu}$ *is a convex combination of rate vectors in* $\mathsf{R}_{L_{MW}}^{\vec{c}}$,
*(i) There exists a point-to-point link* $j \in L_{MW}$ *such that* $\sigma_M \mu_j < \nu_j$, *or*
*(ii) There exists a multiuser link* $(k, l) \in L_{MW}$ *such that* $\sigma_M \mu_k < \nu_k$ *and* $\sigma_M \mu_l < \nu_l$.

The *multiuser local pooling factor* $\sigma_M^*$ of a network is the supremum of all $\sigma_M$ such that every candidate MW subset $L_{MW}$ of $\mathcal{L}$ satisfies $\sigma_M$-local pooling, *i.e.*,

$$\sigma_M^* = \sup \left\{ \sigma_M \mid \forall L_{MW} \in \mathcal{L}, \; \forall \; \vec{c} \in \mathcal{C}_{L_{MW}}, \; \text{conditions (i) or(ii) are satisfied for every } \vec{\mu} \text{ and } \vec{\nu} \right\}, \qquad (6)$$

where $\vec{\mu}$ and $\vec{\nu}$ are again convex combinations of the rate vectors in $\mathsf{R}_{L_{MW}}$ and $\mathsf{R}_{L_{MW}}^{\vec{c}}$ respectively. For the variable rate MGMW scheme, we state the following result that extends the performance analysis results for the MGMW scheme with fixed multiuser link rates.

**Theorem 4.** *For a graph* $\mathcal{G}$ *with the* multiuser local pooling factor $\sigma_M^*$ *as defined in* (6), *variable rate MGMW scheme stabilizes all arrival rate vectors* $\vec{\lambda} \in \sigma_M^* \Lambda$, i.e.,

$$\gamma^* \geq \sigma_M^*. \qquad (7)$$

*Proof:* The proof is given in Appendix D. ∎

We observe that for the variable rate MGMW, the multiuser local pooling conditions involve verifying the conditions over all possible sets of transmission rates associated with the links in $L_{MW}$. This is in contrast to the case where the multiuser link rate is fixed, since there is only transmission rate vector associated with the links in an $L_{MW}$ subset. We now derive an upper bound for the efficiency of variable rate MGMW can further be analyzed

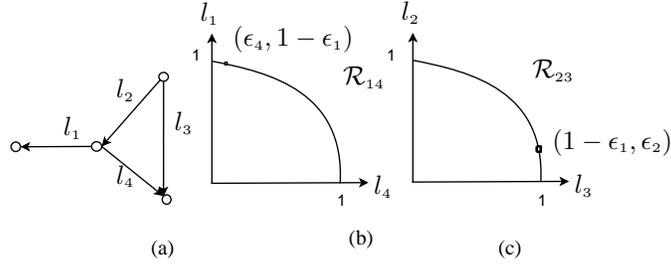

Figure 4: A network with two multiuser links.

as follows. Let $\hat{\sigma}_M$ be defined as:

$$\hat{\sigma}_M = \inf \left\{ \sigma_M \mid \exists\, L_{MW} \in \mathcal{L}, \text{and an associated transmission rate vector } \vec{c} \in \mathcal{C}_{L_{MW}}, \right. \tag{8}$$

$$\left. \text{such that } \sigma_M \vec{\mu} > \vec{\nu} \text{ for some } \vec{\mu},\ \vec{\nu} \right\}, \tag{9}$$

$\vec{\mu}$ and $\vec{\nu}$ being convex combinations of the rate vectors in $\mathsf{R}_{L_{MW}}$ and $\mathsf{R}^x_{L_{MW}}$ respectively.

**Theorem 5.** *For a graph $\mathcal{G}$ with $\hat{\sigma}_M$ as defined in* (9), *there exist arrival rate vectors $\vec{\lambda}$ arbitrarily close to, but outside $\hat{\sigma}_M \Lambda$, such that the network graph is unstable under variable rate MGMW.*

$$\gamma^* \leq \hat{\sigma}_M. \tag{10}$$

*Proof:* The proof is similar to that of the fixed multiuser link rate case and is given in Appendix E. ∎

Theorem 4 and Theorem 5 together imply that $\sigma_M^* \leq \gamma^* \leq \hat{\sigma}_M$. Since the $\sigma_M$-local pooling conditions need to be satisfied by all rate assignments associated with the links that belong to an $L_{MW}$ subset, some of the rate assignments could limit $\hat{\sigma}_M$ more than other rate assignments, leading to a lower value of efficiency for variable rate MGMW.

The stability region of a network with multiuser link becomes even larger when we allow the entire achievable rate region for each multiuser link to be used by the scheduler. However, the proposed variable rate MGMW policy, which is a natural extension of the fixed rate MGMW policy can perform poorly compared to the fixed rate MGMW policy. Indeed, the efficiency ratio of the variable rate MGMW policy is in fact bounded above by the efficiency ratio of a fixed rate MGMW policy (with fixed rates being chosen from the rate region of multiuser links). In the following section, we provide examples of network graphs in which we illustrate how the availability of the entire multiuser capacity rate region can bring down the efficiency of variable rate MGMW. We observe further that GMM is actually throughput optimal for these network graphs examples in the absence of multiuser links. These examples illustrate that it might be more beneficial to use a fixed rate MGMW scheduler whose rates have been carefully chosen from the corresponding rate regions of multiuser links compared to the variable rate MGMW scheduler.

## VI. Performance of variable rate MGMW

In this section, we illustrate the performance of the variable rate MGMW scheme using some examples. Specifically, we provide some cases in which the variable rate MGMW performs much more poorly in terms of the efficiency ratio, compared to the fixed rate MGMW scheduler to the extent that, certain arrival rates that are stabilizable by fixed rate MGMW make the queues blow up for variable rate MGMW. While the optimal network stability region is larger when the rates are allowed to vary, we show that the variable rate MGMW may perform worse relative to fixed rate MGMW with regard to efficiency ratio.

**Example 6.** Consider the network graph shown in Fig. 7a. The network has four edges and is capable of using the multiuser links $(l_1, l_4)$ and $(l_2, l_3)$, whose multiuser capacity regions are given by $\mathcal{R}_{14}$ and $\mathcal{R}_{23}$ respectively. We assume that all the point-to-point links have unit capacities. Let $(1 - \epsilon_1, \epsilon_4) \in \mathcal{R}_{14}$ and $(\epsilon_2, 1 - \epsilon_1) \in \mathcal{R}_{23}$ be two arbitrary rate pairs. We now show that the efficiency of variable rate MGMW is limited by the choice of the multiuser link rate pairs according to the relation:

$$\gamma \leq \frac{1 + \epsilon_1 + \delta}{2}, \tag{11}$$

where $\delta = (\epsilon_4 - \epsilon_1) + (\epsilon_2 - \epsilon_1)$ is a positive quantity that depends on the value of the link rate pairs.

*Proof:* Consider the candidate MW subset $L_{MW} = \{(l_1, l_4), (l_2, l_3)\}$ with the link rates given by $(1 - \epsilon_1, \epsilon_4) \in \mathcal{R}_{14}$ and $(\epsilon_2, 1 - \epsilon_1) \in \mathcal{R}_{23}$, where $\epsilon_1, \epsilon_4 \in (0, 1)$. Let $\vec{\nu}$ be a convex combination of the rate allocation vectors on $L_{MW}$, with the link rates that yield the maximum weight assumed to be $(1 - \epsilon_1, \epsilon_4)$ and $(\epsilon_2, 1 - \epsilon_1)$. We choose $\vec{\nu}$ to be:

$$\vec{\nu} = 0.5 * [1 - \epsilon_1 \ 0 \ 0 \ \epsilon_4] + 0.5 * [0 \ \epsilon_2 \ 1 - \epsilon_1 \ 0].$$

Let $\vec{\mu}$ be a convex combination of the rate allocation vectors in $R_{L_{MW}}$. We choose $\vec{\mu}$ as:

$$\vec{\mu} = 0.5 K \left( (1 - \epsilon_1) \, [1 \ 0 \ 1 \ 0] + \epsilon_2 \, [0 \ 1 \ 0 \ 0] + \epsilon_4 \, [0 \ 0 \ 0 \ 1] \right).$$

Then, $\vec{\mu} = K\vec{\nu}$, which implies from Theorem 5 that the efficiency of variable rate MGMW cannot be greater than $\frac{1}{K}$. For $\vec{\mu}$ to be a convex combination of the rate allocation vectors, it must satisfy $0.5 K[(1 - \epsilon_1) + \epsilon_2 + \epsilon_4] = 1$, implying that

$$\frac{1}{K} = (1 - \epsilon_1 + \epsilon_2 + \epsilon_4)/2. \tag{12}$$

Since the multiuser capacity region is strictly convex, $1 - \epsilon_1 + \epsilon_4 > 1$, which implies that $\epsilon_4 > \epsilon_1$. Similarly, $1 - \epsilon_1 + \epsilon_2 > 1$, which implies that $\epsilon_2 > \epsilon_1$. Let $\delta = (\epsilon_4 - \epsilon_1) + (\epsilon_2 - \epsilon_1)$, so that $\delta > 0$. Substituting the numerator in (12) with $\delta$, we obtain,

$$\frac{1}{K} = (1 + \epsilon_1 + \delta)/2. \tag{13}$$

∎

Eq. (11) suggests that the efficiency of variable rate MGMW scheduler for the network graph in Fig. 7a is limited by those multiuser rate points which have smaller values of $\epsilon_1$ and $\delta$, as they yield lower values of $\hat{\sigma_M}$. In this example, a smaller $\epsilon_1$ and $\delta$ corresponds to those multiuser rate points which are close to one of the edges' point-to-point capacities, as shown in Fig. 7b and 7c.

**Example 7.** Consider the network graph in Fig. 5 with four edges having multiuser links $(1, 4)$ and $(2, 3)$, where $\mathcal{R}_{14}$ and $\mathcal{R}_{23}$ are their respective multiuser capacity regions. Let $(1 - \epsilon_1, \epsilon_4) \in \mathcal{R}_{14}$ and $(1 - \epsilon_2, \epsilon_3) \in \mathcal{R}_{23}$ be two arbitrary rate pairs. Then, we show that the efficiency of variable rate MGMW is upper bounded as :

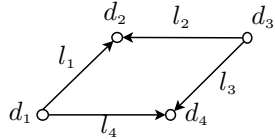

Figure 5: Four edge network with two multiuser links.

$$\gamma <= \frac{2 - (\epsilon_1 + \epsilon_2)}{2}, \text{ where } \epsilon_1 + \epsilon_2 < 1. \tag{14}$$

*Proof:* We choose $L_{MW} = \{(l_1, l_4), (l_2, l_3)\}$ with links having rates $(1 - \epsilon_1, \epsilon_4) \in \mathcal{R}_{14}$ and $(1 - \epsilon_2, \epsilon_3) \in \mathcal{R}_{23}$. Let $\vec{\nu} = 0.5 * [1 - \epsilon_1 \ 0 \ 0 \ \epsilon_4] + 0.5 * [0 \ 1 - \epsilon_2 \ \epsilon_3 \ 0]$ be a convex combination of rate allocation vectors in $L_{MW}$ with the assumed link rates, and $\vec{\mu} = 0.5 * K * (1 - \epsilon_1) * [1 \ 0 \ 1 \ 0] + 0.5 * K * (1 - \epsilon_2) * [0 \ 1 \ 0 \ 1]$ be a convex combination of rate allocation vectors in $\mathsf{R}_{L_{MW}}$. Using the fact that $\vec{\mu}$ is a convex combination, we obtain the relation:

$$\frac{1}{K} = (2 - (\epsilon_1 + \epsilon_2)) / 2. \tag{15}$$

In this case, the condition $\vec{\mu} = K\vec{\nu}$ is satisfied if $\epsilon_1 + \epsilon_3 = 1$, and $\epsilon_2 + \epsilon_4 = 1$. The strict convexity of the multiuser capacity region gives us the conditions $\epsilon_4 > \epsilon_1$, and $\epsilon_3 > \epsilon_2$. Both these conditions together imply that $\epsilon_1 + \epsilon_2 < 1$. ∎

From Eq. (14), we observe that increasing the quantity $\epsilon_1 + \epsilon_2$ decreases the upper bound on the efficiency of variable rate MGMW to close to 0.5, and hence rate pairs in which at least one of the edges get high rates, dominate the performance of variable rate MGMW. Note that in the absence of multiuser links, GMS is throughput optimal for both network graphs [3].

The above examples highlight certain scenarios where the variable rate MGMW scheduler can perform poorly in terms of efficiency ratio, when compared to a MGMW policy whose links operate a fixed rate chosen carefully from the multiuser capacity region. Choosing a fixed rate point also reduces the coding complexity of the multiuser links by requiring fewer number of codebooks.

## VII. Simulation of the Performance of MGMW

In this section, we simulate the performance of MGMW with fixed multiuser link rates and GMM in a randomly connected network graph. Figure 6a shows an arbitrary network graph having point-to-point as well as multiuser links. The multiuser (BC) links in this graph are links (1,2), (4,7), (3,8), (5,6), and (9,14). In this example we chose the transmission rates of the point-to-point links at random, uniformly between 3 and 10 units. The rates of the multiuser links are chosen to ensure the convexity of the rate region is assured. The arrival process for each edge is Bernoulli and we denote the arrival rate with $\lambda_l$.

In Fig. 6b, we plot the total queue size (sum of queue lengths at each edge) as we increase the arrival rate in edges 9 and 14, from 2 to 2.5 as we keep $\lambda_l = 1$ for all other links. Here, the transmission rates of links 9, 14 and (9,14) are 6, 4 and (4,3) respectively. The graph suggests that MGMW yields a constant gain in arrival rate for the multi-user links as each edge of the multiuser link (9,14) is seen to sustain around 5% more traffic, as compared to the GMM case.

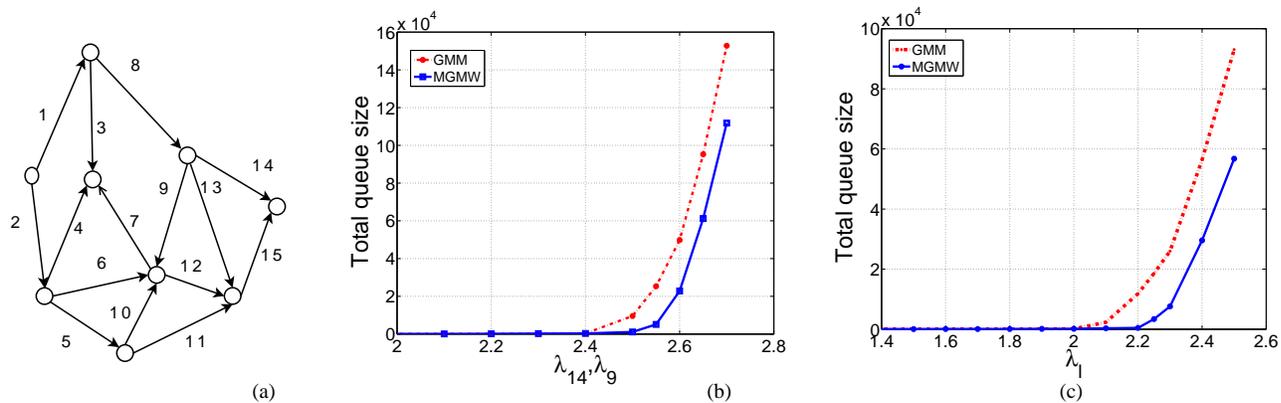

Figure 6: An Arbitrary network graph with 15 edges and 5 broadcast links.

In Fig. 6c, we simultaneously increase the arrival rate across edges 1, 2, 9, 14, 3 and 8 while keeping the arrival rates at other edges fixed at 1. Here, the capacities of these links are 10, 8, 6, 4, 12 and 8, and the transmission rates of the BCs $(1,2)$, $(9,14)$ and $(3,8)$ are (9,5), (4,3) and (10,6) respectively. Similar to the previous scenario, the plots again show that the total queue size with MGMW is lower than that with GMM. Here, each edge of the multiuser links is able to sustain 10% more traffic than in the no multiuser link case. Thus, for the network in Fig. 6a, MGMW appears to stabilize a larger range of arrival rates.

## VIII. CONCLUSIONS

In this paper we explored the problem of link scheduling in a setting that allows for the use of techniques from multi-user information theory. To this end, we proposed a modified version of the binary interference model by introducing the notion of a secondary interference set for each link of the network. The interference model proposed in this paper could be thought of in a loose sense as a hybrid of the binary interference model and the SINR model. Since the optimal algorithm is known to have high complexity (NP hard in many cases), we provided a suboptimal greedy algorithm called MGMW for our interference model. We characterized the performance of MGMW by deriving local pooling conditions and relating the multiuser local pooling factor to the efficiency of MGMW. For a network with capacity region $\Lambda$ and a multiuser local pooling factor $\sigma_M$, we showed that MGMW stabilizes every arrival rate vector in $\sigma_M \Lambda$ and that there exists a non-stabilizable arrival rate vector, arbitrarily close to, but strictly outside of $\sigma_M \Lambda$. We gave examples of certain network graphs where MGMW was throughput optimal and a graph where the multiuser local pooling factor is less than one. We also considered a network model where the scheduler has the freedom to select the multiuser link rate every time slot. We analyzed the performance of the variable rate MGMW scheduler and showed that the availability of the entire rate region could hurt the performance of variable rate MGMW. Finally, we also observed the performance of MGMW in an arbitrary graph and compared it to that of GMM.

# APPENDIX A

## PROOF OF THEOREM 1

The proof shows stability of the network under MGMW by finding a Lyapunov function and showing that it has a negative drift for the fluid limit of the system. The idea of the proof is similar to the stability proof in [6], which is for the scenario of no multiuser links. We assume that the arrival process for each link satisfies conditions for the fluid limit to exist, which is that the Strong Law of Large Numbers (SLLN) should hold for the arrival process. For example, SLLN holds when the packet arrivals in each queue are IID with bounded second moments. Here, we assume a modified arrival process, where we relax the IID assumption in the first time slot alone, and allow a deterministic but finite number of packets to arrive in the first time slot. Note that this does not affect the applicability of SLLN and $\sum_{n=1}^{\infty} A_i(n)/n \to E(A_i(n)) = \lambda_i$w.p 1. Let $\vec{A(t)}$ denote the cumulative arrival process and $\vec{S(t)}$ denote the cumulative service process until time slot $t$. For the arrival and service processes, we use $A_l(t) = A_l(\lfloor t \rfloor)$, and $S_l(t) = S_l(\lfloor t \rfloor)$. For the queue process $Q_l(t)$, we employ linear interpolation. We now consider a sequence of scaled queuing systems $(\vec{Q^n}(\cdot), \vec{A^n}(\cdot), \vec{S^n}(\cdot))$. where we apply the scaling $Q_l^n(nt)/n$, $A_l(nt)/n$, and $S_l(nt)/n$, $\forall l \in \mathcal{E}$ with the queue process satisfying $\sum_{l \in \mathcal{E}} Q_l^n(0) \leq n$. Then, using the techniques to establish fluid limit in [11], one can show that a fluid limit exists almost surely, *i.e.* for almost all sample paths and for any positive $n \to \infty$, there exists a sub-sequence $n_j$ with $n_j \to \infty$ such that following convergence holds uniformly over compact sets. For all $l \in \mathcal{E}$, $\frac{1}{n_j} A_l^{n_j}(n_j t) \to \lambda_l t$, $\frac{1}{n_j} S_l^{n_j}(t) \to s_l(t)$, and $\frac{1}{n_j} Q_l^{n_j}(n_j t) \to q_l(t)$. $q_l(t)$ and $s_l(t)$ are the fluid limits for the queue length processes and the service rate processes respectively. The fluid limit is absolutely continuous and hence the derivative of $q_l(t)$ exists almost everywhere [11] satisfying:

$$\frac{d}{dt} q_l(t) = \begin{cases} [\lambda_l - \pi_l(t)]^+ & q_l(t) > 0. \\ 0 & \text{otherwise.} \end{cases} \tag{16}$$

where $\pi_l(t) = \frac{d}{dt}(s_l(t))$. We now show that the largest queue weighted rate (taken over point-to-point link or broadcast links) of the fluid limit model always decreases under the MGMW algorithm. This allows us to define the Lyapunov function for the system as the maximum weight over all links and establish its drift to be negative. Consider the times $t$ when the derivative $\frac{d}{dt} q_l(t)$ exists for all $l \in \mathcal{E}$. Let $L_0(t)$ denote the set of links with the largest weight, *i.e.*,

$$L_0(t) = \{m \in \mathcal{L} \mid w_m = \max_{m \in \mathcal{L}} w_m\}.$$

Define the derivative of the weights of links in $\mathcal{L}$ as follows:

$$\hat{w}_m(t) = \begin{cases} \frac{d}{dt} q_j(t) c_j & m \text{ is a point-to-point link } j \\ \frac{d}{dt} q_k(t) c_{kl} + q_l(t) c_{lk} & m \text{ is a} \\ & \text{multiuser link } (k, l) \end{cases} .$$

Let $L(t)$ denote the set of links from $L_0(t)$, which have the maximum derivative of the weights,

$$L(t) = \{m \in \mathcal{L} \mid \hat{w}_m(t) = \max_{m \in L_0(t)} \hat{w}_m(t)\}.$$

Then, one can find a small $\delta$ such that in the interval $(t, t + \delta)$, links in $L(t)$ will have the highest weights in that time interval, *i.e.*, $\min_{m \in L(t)} \hat{w}_m(\tau) > \max_{m \in \mathcal{L} \setminus L(t)} \hat{w}(\tau)$ for all $\tau$ in $(t, t + \delta)$. MGMW will select links from the set $L(t)$ first in the interval $(t, t + \delta)$, since it picks the links in decreasing order of weights. If we focus on the links in $L(t)$, then any rate allocation vector selected by MGMW in $(t, t + \delta)$, projected on the set $L(t)$ would yield a rate allocation vector that is an element of $\mathsf{R}_{L(t)}$. This in turn implies that $\vec{\pi}(t)$, the service rate vector under MGMW projected on $L(t)$ is a convex combination of the elements of $\tilde{\mathsf{R}}_{L(t)}$. A formal argument to show this is mostly identical to that in [6] and is omitted here. Because of the convexity condition $\frac{c_{kl}}{c_k} + \frac{c_{lk}}{c_l} > 1$, $L(t)$ is a candidate MW subset. Consider a $\vec{\lambda}$ lying strictly within $\sigma_M^L \Lambda$. Since $\sigma_M^L$ is the local pooling factor, and $L(t)$ is a candidate MW set, it follows from the multiuser local pooling conditions that there exists a link $m \in L(t)$ that satisfies

$$\lambda_j < \pi_j(t), \text{ if } m \text{ is point-to-point link } j, \tag{17}$$

$$\lambda_k c_{kl} + \lambda_l c_{lk} < \pi_k(t)c_{kl} + \pi_l(t)c_{lk}, \text{ if } m \text{ is multiuser link } (k, l).$$

For any candidate MW set $L_{MW}$, let $\vec{s}_{L_{MW}}$ be any convex combination of the elements (rate allocation vectors) of the set $\tilde{R}_{L_{MW}}$. Since every $L_{MW} \subset \mathcal{L}$ satisfies $\sigma_M^L$ local pooling, and $\vec{\lambda}$ lies strictly within $\sigma_M^L \Lambda$, the quantity

$$\epsilon_{\vec{s}_{L_{MW}}} = \max_{\substack{(k,l) \in L_{MW} \\ j \in L_{MW}}} \left( (s_j - \lambda_j), (s_k c_{kl} + s_l c_{lk} - \lambda_k c_{kl} - \lambda_l c_{lk}) \right)$$

is strictly positive for every $L_{MW}$. We then define the infimum of all such positive quantities over all such subsets $L_{MW}$ and all vectors $\vec{s}_{L_{MW}}$ as

$$\epsilon^* = \inf_{\{\vec{s}_{L_{MW}} \forall L_{MW} \subset \mathcal{L}\}} \epsilon_{\vec{s}_{L_{MW}}}. \tag{18}$$

and we observe that $\epsilon^* > 0$. Hence, from the relation in (17), there exists $m \in L(t)$ such that the following holds:

$$\lambda_j - \pi_j(t) \le -\epsilon^*, \text{ if } m \text{ is point-to-point link } j, \tag{19}$$

$$(\lambda_k - \pi_k(t))c_{kl} + (\lambda_l - \pi_l(t))c_{lk} \le -\epsilon^*, \text{ if } m \text{ is a link } (k, l).$$

From (16), and the fact that all links in $L(t)$ have the same derivative, (19) implies that $\hat{w}_m \le -\epsilon^*, \ \forall m \in L(t)$. Hence, we observe that there exists no link in $L(t)$ with $\hat{w}_m \ge 0$, and $q_m > 0$, where $q_m = \max(q_k, q_l)$, if m is a multiuser link $(k, l)$. Now, we can consider the following Lyapunov function $V(t) := \max_{m \in \mathcal{L}} w_m$. For $V(t) > 0$, we have that

$$\frac{d^+}{dt^+} V(t) \le \max_{m \in \mathcal{L}} \hat{w}_m \le -\epsilon^*. \tag{20}$$

where $\frac{d^+}{dt^+} V(t) = \lim_{\delta \downarrow 0} \frac{V(t+\delta) - V(t)}{\delta}$ is the right hand derivative of $V(t)$. This implies that the largest weight must decrease in the time interval $(t, t+\delta)$. Since the above inequality holds almost everywhere in $t$, the negative drift of the Lyapunov function implies that the fluid limit model of the system is stable and hence by the result in [11], the original system is also stable.

## Appendix B

### Proof of Theorem 2

We construct an arrival traffic using the approach in [6] and show that under this traffic pattern the network is unstable under MGMW. Let $J$ denote the number of possible rate allocations on the set $L_{MW}$. The vector $\vec{\nu}$, being a convex combination of the elements in $\tilde{R}_{L_{MW}}$, can be written as

$$\vec{\nu} = \sum_{i=0}^{J-1} \omega_i \vec{r}_i, \qquad \vec{r}_i \in \tilde{R}_{L_{MW}}$$

where $\omega_i \ge 0$ for all $0 \le i \le J-1$ and $\sum_{i=0}^{J-1} \omega_i = 1$. Let $v_i$ be a rational number which satisfies $\sum_{i=0}^{J-1} |\omega_i - v_i| \le \frac{\delta}{J}$ for any $\delta > 0$. Such a $v_i$ clearly exists for every $\omega_i$. To enable the construction of the traffic pattern we now define a new vector $\hat{\vec{\nu}} = \sum_{i=0}^{J-1} v_i \vec{r}_i$. Thus one can make the vector $\hat{\vec{\nu}}$ arbitrarily close to $\vec{\nu}$. We now specify our arrival traffic with load $\hat{\vec{\nu}} + \epsilon \vec{k}$, such that the system is unstable under MGMW. The arrival traffic for every queue consists of IID packet arrivals in each time slot, except for the first time slot alone. Packets arrive at the beginning of time slots. Without loss of generality we assume that for the candidate MW subset $L_{MW}$, if $(k, l)$ is a multiuser link in $L_{MW}$, then $k$ denotes the edge of the link $(k, l)$ that can be present as a point-to-point link in $L_{MW}$. Let the initial queue state vector be $\vec{Q_0} \succeq 0$. Let $j$ denote any point-to-point link belonging to $L_{MW}$, and $(k, l)$ denote any multiuser link in $L_{MW}$. Then the packet arrivals in the first time slot are such that $\vec{Q_0}$ satisfies the following constraints:

$$Q_j c_j = Q_k c_{kl} + Q_l c_{lk} = Q_k c_k + c_k^2 - c_k c_{kl} > Q_l c_l + c_l c_{lk}, \qquad (21)$$

for all links $j$, $(k, l) \in L_{MW}$. We show in Lemma 1 that a vector $\vec{Q_0} \succeq 0$ satisfying (21) exists. We now describe the statistics of our arrival traffic. Let vector $\vec{r_i}$ be picked with probability $\nu_i$. Conditioned on $\vec{r_i}$ being picked, one of two events may occur:

**1.** With probability $1 - \epsilon$,

(a) $r_i(j)$ packets arrive into every point-to-point link $j \in L_{MW}$ such that $r_i(j) = c_j > 0$, and

(b) $c_{kl}$ and $c_{lk}$ packets arrive into links $k$ and $l$ respectively $\forall (k, l) \in L_{MW}$ such that $r_i(k) > 0$ and $r_i(l) > 0$.

We show that when packets arrive in the manner described in event **1**, MGMW picks rate allocation vector $\vec{r_i}$, and at the end of time slot $t$, the queues in $L_{MW}$ continue to satisfy the relation in (21).

(i) The weight of any point-to-point link $j \in L_{MW}$ satisying $r_i(j) = c_j$, is given by $Q_j c_j + c_j^2$. Using (21) we obtain:

$$Q_j c_j + c_j^2 > Q_l c_l, \ \forall l \in L_{MW} \text{ such that } r_i(l) = 0, \text{ and}$$

$$Q_j c_j + c_j^2 > Q_m c_{mn} + Q_n c_{nm} > Q_m c_m,$$

$$\forall (m, n) \in L_{MW} \text{ such that } r_i(m) = r_i(n) = 0. \qquad (22)$$

The inequality in (22) follows since $c_m^2 - c_m c_{mn} > 0$.

(ii) For any multiuser link $(k, l)$ satisfying $r_i(k) = c_{kl}$ and $r_i(k) = c_{kl}$,

$$(Q_k + c_{kl})c_{kl} + (Q_l + c_{lk})c_{lk} > Q_j c_j, \ \forall j \in L_{MW} \text{ s. t } r_i(j) = 0,$$

$$Q_k c_{kl} + Q_l c_{lk} + c_{kl}^2 + c_{lk}^2 > Q_m c_{mn} + Q_n c_{nm},$$

$$\forall (m, n) \in L_{MW} \text{ such that } r_i(m) = r_i(n) = 0,$$

$$Q_k c_{kl} + Q_l c_{lk} + c_{kl}^2 + c_{lk}^2 > Q_l c_l + c_l c_{lk},$$

$$Q_k c_{kl} + Q_l c_{lk} + c_{kl}^2 + c_{lk}^2 > Q_k c_k + c_k c_{kl}, \qquad (23)$$

where (23) holds since

$$Q_k c_{kl} + Q_l c_{lk} + c_{kl}^2 + c_{lk}^2 = Q_k c_k + c_k c_{kl} + (c_k - c_{kl})^2 + c_{lk}^2.$$

(iii) For link $k \in L_{MW}$ satisfying $r_i(k) = c_k$ and $(k, l) \in L_{MW}$ for some $l$, we have:

$$Q_k c_k + c_k^2 > Q_j c_j, \ \forall j \in L_{MW} \text{ with } r_i(j) = 0$$

$$Q_k c_k + c_k^2 > Q_m c_{mn} + Q_n c_{nm} > Q_m c_m,$$

$$Q_k c_k + c_k^2 > Q_l c_l, \text{ and}$$

$$Q_k c_k + c_k^2 = (Q_k + c_k)c_{kl} + Q_l c_{lk}. \qquad (24)$$

The weight of link $k$ therefore dominates all the links not served by $\vec{r}_i$. MGMW breaks the tie between the weights of $k$ and $(k, l)$ shown in (24) by scheduling link $k$. (i), (ii) and (iii) establish that weight of links served in $\vec{r}_i$ dominate the weight of other links and hence MGMW schedules $\vec{r}_i$.

**2.** With probability $\epsilon$, packets arrive into the queues in $L_{MW}$ in the following manner.

(a) $c_j + \hat{c}_j$ packets arrive into point-to-point links $j \in L_{MW}$ for which $r_i(j) = c_j$.

(b) $c_{kl} + \hat{c}_k$ packets into link $k$ and $c_{lk} + \hat{c}_l$ packets arrive into the queues of $k$ and $l$ for multiuser links $(k, l)$ such that $r_i(k) > 0$ and $r_i(l) > 0$.

(c) $\hat{c}_m$ packets arrive into all other links $m \in L_{MW}$ for which $r_i(m) = 0$.

The quantities $\hat{c}_j$, $\hat{c}_k$, and $\hat{c}_l$ are such that they satisfy the following weight criteria:

$$\hat{c}_j c_j = \hat{c}_k c_k = \hat{c}_k c_{kl} + \hat{c}_l c_{lk} > \hat{c}_l c_l. \tag{25}$$

It can be shown using an argument identical to that used for (21) that there exist positive $\hat{c}_j$, $\hat{c}_k$, and $\hat{c}_l$ that satisfy relation (25). When packets arrive according to the event **2**, MGMW still schedules $\vec{r}_i$. This is because (21) and (25) yield

$$(Q_j + \hat{c}_j)c_j = (Q_k + \hat{c}_k)c_{kl} + (Q_l + \hat{c}_l)c_{lk}$$

$$= (Q_k + \hat{c}_k)c_k + c_k^2 - c_k c_{kl}$$

$$> (Q_l + \hat{c}_l)c_l + c_l c_{lk}.$$

which again satisfies relation (21). However, at the end of time slot $t$, when packets arrive as in event **2**, the length of each queue $j \in \mathcal{E}_{L_{MW}}$ increases by a fixed positive quantity $\hat{c}_j$. We can now describe the queue evolution for our arrival traffic. The initial queue state satisfies (21). Therefore at the end of each time slot, with probability $1 - \epsilon$, the queues of all edges $j \in \mathcal{E}_{L_{MW}}$ remain unchanged and with probability $\epsilon$, the queues increase by a fixed positive quantity. Since the queues in $L_{MW}$ are non-decreasing, and the event that the queue length increases by a fixed positive quantity occurs infinitely often, the system is unstable under the MGMW scheme. The arrival rate of our proposed arrival traffic is determined as follows. Let $\vec{k}$ denote the vector defined as:

$$\vec{k}(j) = \begin{cases} \hat{c}_j & j \in \mathcal{E}_{L_{MW}} \\ 0 & j \notin \mathcal{E}_{L_{MW}}. \end{cases}$$

The arrival rate is then given by :

$$\vec{\lambda} = \sum_{i=0}^{J-1} (v_i(1 - \epsilon)\vec{r}_i + v_i \epsilon(\vec{r}_i + \vec{k})) = \vec{\nu} + \epsilon \vec{k}.$$

.

**Lemma 1.** *There exist queue lengths $Q_1, Q_2, \ldots Q_M$, where $M$ is the number of edges in $L_{MW}$, such that they satisfy:*

$$Q_j c_j \overset{1}{=} Q_k c_{kl} + Q_l c_{lk} \overset{2}{=} Q_k c_k + c_k^2 - c_k c_{kl} \overset{3}{>} Q_l c_l + c_l c_{lk}. \tag{26}$$

*Proof:* Consider $K$ such that

$$Q_j = \frac{K}{c_j}, \; Q_k = \frac{K + c_k c_{kl} - c_k^2}{c_k}. \tag{27}$$

Then, to satisfy equality 2, we need to choose

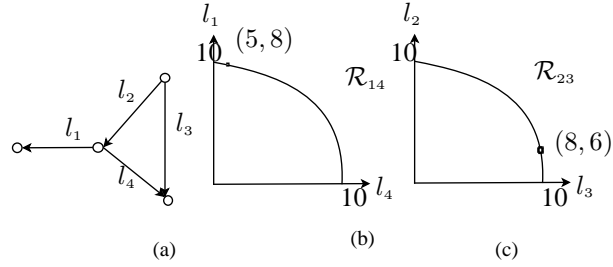

Figure 7: A network with two multiuser links.

$$Q_l = \frac{Q_k(c_k - c_{kl}) + c_k^2 - c_k c_{kl}}{c_{lk}} \tag{28}$$

Now, for inequality (3) to hold, substituting for $Q_l$ we have,

$$Q_k \frac{(c_k c_{lk} + c_l c_{kl} - c_l c_k)}{c_{lk}} > \frac{c_l}{c_{lk}} \left[ c_k^2 - c_k c_{kl} \right] +$$

$$c_l c_{lk} - c_k^2 + c_k c_{kl}. \tag{29}$$

From our assumption that $c_{kl}/c_k + c_{lk}/c_l > 1$, the left hand side of (29) is positive. Hence for (29) to hold, we need

$$Q_k > \left[ \frac{(c_k^2 - c_k c_{kl})(c_l - c_{lk}) + c_l c_{lk}^2}{c_l c_{kl} + c_k c_{lk} - c_l c_k} \right].$$

This must be satisfied for every $(k, l) \in L_{MW}$.

$$\text{Let } p = \max_{(k,l) \in L_{MW}} \left[ \frac{(c_k^2 - c_k c_{kl})(c_l - c_{lk}) + c_l c_{lk}^2}{c_l c_{kl} + c_k c_{lk} - c_l c_k} \right].$$

Then choose $K$ such that $(K + c_k c_{kl} - c_k^2)/c_k > p$ for all links $(k, l) \in L_{MW}$. Then, (27) gives the values of $Q_j$ and $Q_k$. The value of $Q_l$ corresponding to $Q_k$ is then obtained from (28). Thus one can find a $K > 0$ and hence a positive queue vector $\vec{Q}_0$ satisfying relation (21). In a similar manner one can find positive $\hat{c}_j$, $\hat{c}_k$, and $\hat{c}_l$ that satisfy relation (25). ∎

In the example given below, we find the initial queue state vector $\vec{Q}_0$ for a given $L_{MW}$. To illustrate the proof, we provide the following simulation example:

We consider the network graph described in Example 6. The individual link capacities are 10 packets, while the fixed rate for the BC links (1,4) and (2,3) are (5,8) and (8,5) respectively. Considering $L_{MW}$ to be the set $\{(1, 4), (2, 3)\}$, and by choosing $\vec{\mu} = 0.5 * 20/19 * (0.8\, [10\ 0\ 10\ 0] + 0.6\, [0\ 10\ 0\ 0] + 0.5\, [0\ 0\ 0\ 10])$ and $\vec{v} = 0.5 * [8\ 0\ 0\ 6] + 0.5 * [0\ 5\ 8\ 0]$, we obtain $\sigma_M^U = 19/20$, implying that the efficiency $\gamma \le 19/20$ for this network graph. We then simulate the unstable traffic pattern described in the proof of Theorem 2 and plot the queue sizes in Fig. B.

The figure shows the queue sizes increasing over the observed time slots, thus verifying Theorem 2 for Example 6. Packets arrive according to rate allocation vector $[8\ 0\ 0\ 6]$ with probability 0.5, and the vector $[5\ 0\ 0\ 8]$ with probability 0.5. Additional packets may arrive into the queues with probability 0.0002. The additional packets are such that they satisfy the constraint in eq.(24), in the proof of Theorem 2.

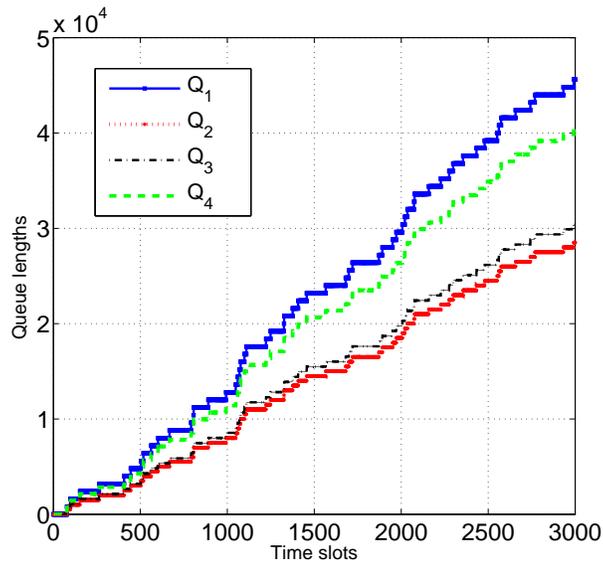

Figure 8: Plot of queue sizes for the links in Fig. 7a

# Appendix C

## Proof of theorem 3

The proof idea can be summarised as follows: Every candidate MW subset $L_{MW}$ of a tree network graph $\mathcal{G}$ has a node $d_0$, such that every rate allocation vector on $L_{MW}$ schedules one of the links connected to $d_0$. Any two convex combinations of rate vectors will therefore satisfy the multiuser local pooling conditions for one of these links. We prove this by using a linear program to represent the multiuser local pooling conditions.

**Lemma 2.** *For the candidate MW subset $L_{MW}$, let $\sigma_M^{L_{MW}}$ denote the highest value of $\sigma_M$ for which $L_{MW}$ satisfies the multiuser local pooling conditions specified in Definition 3. $\mathsf{M}$ is defined as a matrix whose columns consist of all rate allocation vectors in the set $\mathsf{R}_{L_{MW}}$. $\tilde{\mathsf{M}}$ denotes the matrix whose columns are all the rate allocation vectors in the set $\tilde{\mathsf{R}}_{L_{MW}}$. Also, let $\vec{e}$ and $\vec{\tilde{e}}$ denote all ones column vectors of length $|\mathsf{R}_{L_{MW}}|$ and $|\tilde{\mathsf{R}}_{L_{MW}}|$ respectively.*

$\sigma_M^{L_{MW}}$ *can then be represented as the solution to the linear program given below:*

$$\max_{\vec{x} \geq 0} \tau \tag{30}$$

*subject to* $\vec{x}\mathsf{M}^0 \preceq \vec{e}$

$\vec{x}\tilde{\mathsf{M}}^0 \succeq \tau.$

$x_k = 0, \ if \ (k,l) \in L_{MW} \ and \ k \notin L_{MW},$

*where*

$$\tilde{\mathsf{M}}^0_{l_{ij}} = \begin{cases} \tilde{\mathsf{M}}_{kj}c_{kl} + \tilde{\mathsf{M}}_{lj}c_{lk}, & if \ (k,l) \in L_{MW}, \\ \tilde{\mathsf{M}}_{ij} & otherwise. \end{cases}$$

$$\mathsf{M}^0_{l_{ij}} = \begin{cases} \mathsf{M}_{kj}c_{kl} + \mathsf{M}_{ij}c_{lk}, & if \ (k,l) \in L_{MW}, \\ \mathsf{M}_{ij} & otherwise. \end{cases}$$

*where* $\mathsf{M}_{ij}$ *denotes the index* $(i,j)$ *of the matrix* $\mathsf{M}$.

*Proof:* $\sigma_M^{L_{MW}}$ can be written as the solution to the following linear program which is obtained from the multiuser local pooling conditions.

$$\inf_{\sigma, \vec{\alpha}, \vec{\beta} \geq 0} \sigma_M \tag{31}$$

subject to $\sigma \left(\mathsf{M}\vec{\alpha}\right)_j \geqslant (\tilde{\mathsf{M}}\vec{\beta})_j$, for point-to-point link $j$,

$\sigma \left(\mathsf{M}\vec{\alpha}\right)_k c_{kl} + \left(\mathsf{M}\vec{\alpha}\right)_l c_{lk} \geqslant (\tilde{\mathsf{M}}\vec{\beta})_k c_{kl} + (\tilde{\mathsf{M}}\vec{\beta})_l c_{lk}$

for multiuser links $(k,l)$,

and $\vec{\alpha}'\vec{e} = 1, \ \vec{\beta}'\vec{e} = 1, \vec{\alpha} \geq 0, \ \vec{\beta} \succeq 0.$

Here $\left(\mathsf{M}\vec{\alpha}\right)_l$ denotes index $l$ of the vector $\mathsf{M}\vec{\alpha}$. Setting $\gamma = \sigma\vec{\alpha}$, and denoting $l_k$ as the edge paired with $k$ in the multiuser link $(k,l)$, the dual function of the linear program is given by:

$$\inf_{\sigma, \vec{\gamma} \succeq 0, \vec{\beta} \succeq 0} \left\{ \sigma + \sum_{i=1}^{n_1} x_i[(\tilde{\mathsf{M}}\vec{\beta})_i - (\mathsf{M}\vec{\gamma})_i] + \right.$$

$$\sum_{k=1}^{n_2} x_k[(\tilde{\mathsf{M}}\vec{\beta})_k c_{kl_k} + (\tilde{\mathsf{M}}\vec{\beta})_{l_k} c_{l_k k} - (\mathsf{M}\vec{\gamma})_k c_{kl_k} - (\mathsf{M}\vec{\gamma})_{l_k} c_{l_k k}]$$

$$\left. + y(\vec{\gamma}'\vec{e} - \sigma) + z(\vec{\beta}'\vec{e} - 1) \right\}.$$

Since

$$(\mathsf{M}\vec{\gamma})_i = \sum_{j=1}^{|\mathsf{R}_{L_{MW}}|} (\mathsf{M}_{ij})\gamma_i \ \text{and} \ (\tilde{\mathsf{M}}\vec{\gamma})_i = \sum_{j=1}^{|\tilde{\mathsf{R}}_{L_{MW}}|} (\tilde{\mathsf{M}}_{ij})\gamma_i,$$

the dual problem can be expressed as

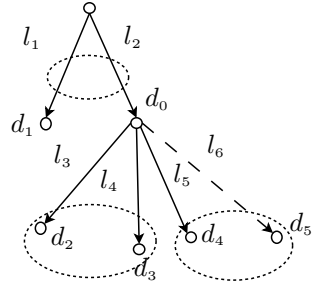

Figure 9: Example of a possible $L_{MW}$ for a tree network. Here $L_{MW} = \{l_2, (l_1, l_2), (l_3, l_4), l_6, (l_5, l_6)\}$. The multiuser links are the pairs of edges enclosed by ellipses in the figure.

$$\max_{\vec{x} \succeq 0, y, z} \inf_{\sigma, \vec{\gamma}, \vec{\beta}} \left\{ \sigma(1-y) - z + y\vec{e}'\vec{\gamma} - \sum_{j=1}^{|\mathsf{R}_{L_{MW}}|} \sum_{i=1}^{n_1} x_i \mathsf{M}_{ij} \gamma_j \right.$$
$$- \sum_{j=1}^{|\mathsf{R}_{L_{MW}}|} \left( \sum_{k=1}^{n_2} x_k [(\mathsf{M}_{k_j} c_{k l_k} + \mathsf{M}_{i_{k_j}} c_{l_k k})] \right) \gamma_j$$
$$\left. + z\vec{\beta}'\vec{e} + \sum_{j=1}^{|\tilde{\mathsf{R}}_{L_{MW}}|} \left( \sum_{k=1}^{n_2} x_k [(\tilde{\mathsf{M}}_{j_k} c_{k l_k} + \tilde{\mathsf{M}}_{j l_k} c_{l_k k})] \right) \beta_j \right\}.$$

The above maximization can then be reduced to

$$\max_{\vec{x} \succeq 0, z} -z$$

subject to

$$y\left( e_j - \sum_{i=1}^{n_1} x_i \mathsf{M}_{ij} - \right.$$
$$\left. \sum_{k=1}^{n_2} x_k [(\mathsf{M}_{k_j} c_{k l_k} + \mathsf{M}_{i_{k_j}} c_{l_k k})] \right) \geq 0, \ j = 1, \ldots, |\mathsf{R}_{L_{MW}}|$$
$$z e_j + \sum_{i=1}^{n_1} x_i \tilde{\mathsf{M}}_{ij} +$$
$$\sum_{k=1}^{n_2} x_k [(\tilde{\mathsf{M}}_{k_j} c_{k l_k} + \tilde{\mathsf{M}}_{i_{k_j}} c_{l_k k})] \geq 0, \ j = 1, \ldots, |\tilde{\mathsf{R}}_{L_{MW}}|,$$
$$y = 1.$$

The dual problem in (30) then follows by defining $\tau = -z$, and extending the length of $\vec{x}$ to $|\mathcal{E}_{L_{MW}}|$ so that it has $|\mathcal{E}_{L_{MW}}| - n_1 - n_2$ zero elements. ∎

The dual problem is to find an $\vec{x} \succeq 0$ that maximizes the value of $\tau$ for which the constraints in (30) are satisfied. From the dual problem, the set $L_{MW}$ satisfies 1-local pooling if $\tau = 1$ is a solution to the problem, which implies that there exists an $\vec{x}$, such that constraint (30) is satisfied with equality.

We now show that for every $L_{MW}$, one can find an $\vec{x}$ to satisfy the equality constraint in (30). Since $L_{MW}$ is a set of links from a tree network graph, it satisfies one or both of the following conditions:

(1) $L_{MW}$ has an isolated point-to-point link, *i.e.,* it consists of two nodes of degree one each that are only connected

to each other.

(2) $L_{MW}$ has at least one node with degree 1.

If $L_{MW}$ has an isolated point-to-point link, let us denote the link by $l$. We then set the $l^{th}$ index of the vector $\vec{x}$, corresponding to link $l$ to be one and all other elements of $\vec{x}$ to be zero. Since link $l$, being an isolated link is always served under the node exclusive interference model, every column vector of $M_{L_{MW}}$ has its $l^{th}$ index equal to one. Such an $\vec{x}$ yields $\tau = 1$ for the dual problem in (30), and hence $L_{MW}$ satisfies 1-local pooling.

When $L_{MW}$ satisfies condition 2, we focus on the node to which the node of degree 1 is connected and denote it by $d_0$. An example of $L_{MW}$ and the node $d_0$ is shown in Fig. 9. We now describe the construction of the vector $\vec{x}$. The elements of this vector depends on the type of links in $L_{MW}$ connected to $d_0$. The links can belong to following types:

(1) point-to-point links connected to $d_0$. If $j$ is a point-to-point link connecting $d_0$, then we set $x_j = 1/c_j$.

(2) Multiuser links that have node $d_0$ as transmitter or receiver. If a multiuser link $(m, n) \in L_{MW}$ is such that edges $m$ or $n$ are connected to $d_0$, and if the point-to-point links $m$ and $n$ are not in $L_{MW}$, then we set $x_m = 0$ and $x_n = \frac{2}{c_{mn}^2 + c_{nm}^2}$.

(3) If a multiuser link $(m, n) \in L_{MW}$ is such that edges $m$ or $n$ are connected to $d_0$, and if say point-to-point links $m \in L_{MW}$, then set $x_n = \left(1 - \frac{1}{c_k}\right)/c_{ik}^2$. Note that for each $m$, the assignment is possible for only one element $x_n$ from our assumption that $Y_m \cap Y_n = \{\Phi\} \forall m, n \in \mathcal{E}$.

Finally we set all other indices in $\vec{x}$ to zero. We now show that if $\vec{x}$ is chosen as defined, it satisfies the constraints in (30) for $\tau = 1$. We first note that under the node-exclusive interference model, amongst the links that share $d_0$ as a common node, only one of the links may be active. As a result, every column $j$ of $\tilde{M}^0$ satisfies one of the following conditions:

(i) $\tilde{M}_{ij}^0 = c_i$, for only one point-to-point link $i \in L_{MW}$ that is connected to $d_0$, in which case $\tilde{M}_k^0 = 0$ for any any other link $k \in L_{MW}$, $k \neq j$ having $d_0$ as one of its nodes.

(ii) $\tilde{M}_{mj}^0 = c_m n$, $\tilde{M}_{nj}^0 = c_{mn}^2 + c_{nm}^2$ for a multiuser link $(m, n) \in L_{MW}$ where at least one of edges $m$ or $n$ are connected to $d_0$. In this case $\tilde{M}_{lj}^0 = 0$ for any other edge $l \in \mathcal{E}_{L_{MW}}$, $l \neq m$ that has $d_0$ as one of its nodes.

(iii) $\tilde{M}_{mj}^0 = c_m$, $\tilde{M}_{nj}^0 = c_m c_{mn}$ for a multiuser link, such that $(m, n) \in L_{MW}$ and $n \in L_{MW}$, where $m$ has $d_0$ as one of its nodes. Again, $\tilde{M}_{lj}^0 = 0$ for any other edge $l \in \mathcal{E}_{L_{MW}}$, $l \neq m$ that has $d_0$ as one of its nodes.

Additionally, we observe for $\tilde{M}^0$ that if condition (iii) is satisfied by some column $j$ for some link $(m, n) \in L_{MW}$, then any column that allocates non-zero rates to $m, n$ satisfies condition (iii). This is because MGMW always selects link $m$ as it gives priority to the point-to-point link over the multiuser link. In a similar manner, we can conclude that if for some column $j$, condition (ii) is satisfied by some link $(m, n) \in L_{MW}$ then every column that allocates non-zero rates to link $(m, n)$ satisfies (ii). This is a consequence of the fact that condition (ii) implies that $(m, n) \in L_{MW}$ while $m \notin L_{MW}$, $n \notin L_{MW}$.

We now exploit the structure of $\tilde{M}^0$ to show that $\vec{x}'\tilde{M}^0 = \vec{e}$ is satisfied by the $\vec{x}$ that we have constructed. Consider any column $j$ in $\tilde{M}^0$. If $j$ satisfies condition (i) for some $i$, then $x_i = 1/c_i$ yields the inner product of $\vec{x}$ and column $j$ to be 1. If $j$ satisfies (ii) for some $(m, n$ then $x_m = 0$, $x_n = 1/(c_{mn}^2 + c_{nm}^2)$ again yielding the inner product as one. A similar conclusion holds when $j$ satisfies (iii) for some $(m, n)$ with $x_m = 1/c_m$, $x_n = 0$. Since every column $j$ satisfies one of the three conditions, $\vec{x}'\tilde{\mathsf{M}}^0 = \vec{\tilde{e}}$ is indeed satisfied. It now remains to be verified that $\vec{x}'\mathsf{M}^0 \leq \vec{e}$.

Consider any multiuser link $(m, n) \in L_{MW}$ such that the edges $m$ or $n$ have $d_0$ as a node. Then for any column $j$, $(\mathsf{M}_{mj}^0, \mathsf{M}_{nj}^0)$ may take on the values $(c_m, c_m c_{mn})$, $(0, c_n c_{mn})$, and $(c_m n, c_{mn}^2 + c_{nm}^2)$. When $x_m = 1/c_m$, $x_n = 0$, the inner product is less than or equal to one. If $x_m = 0$, $x_n = 1/(c_{mn}^2 + c_{nm}^2)$, then the inner product is less than or equal to one since $c_{mn}^2 + c_{nm}^2 > max(c_n c_{nm}, c_m c_{mn})$. If column $j$ has a non zero rate for a point-to-point link connected to $d_0$, the inner product is equal to one.

our construction of the vector $\vec{x}$ ensures that the constraints in (30) are satisfied for $\tau = 1$. This implies that 1-local pooling is satisfied by $L_{MW}$. Since every $L_{MW}$ satisfies 1-local pooling, from Theorem 1, MGMW is throughput optimal for network graph $\mathcal{G}$. $\qquad\square$

As a corollary of 2, we have the following result that yields a lower bound on the value of $\sigma_M^{L_{MW}}$.

$$\sigma_{L_{MW}} \geq \frac{\min_{i \in 1 \ldots |\tilde{\mathsf{R}}_{L_{MW}}|} \|\vec{\tilde{r}^i}\mathbf{T}\|_1}{\max_{j \in 1 \ldots |\mathsf{R}_{L_{MW}}|} \|\vec{r^j}\mathbf{T}\|_1}, \tag{32}$$

This result follows by first setting all the non-zero elements of $\vec{x}$ to 1, i.e, by setting $x_j = 1$, for every $j$ satisfying $j \in L_{MW}$, or $(i, j) \in L_{MW}$ for some multiuser link $(i, j)$. One can then normalize $\vec{x}$ by $\max_{j \in 1 \ldots |\mathsf{R}_{L_{MW}}|} \|\vec{r^j}\mathbf{T}\|_1$ to obtain relation (32).

## APPENDIX D
## PROOF OF THEOREM 4

*Proof:* We first show that if the *multiuser local pooling factor* of a graph $\mathcal{G}(\mathcal{V}, \mathcal{E})$ is $\sigma_M^*$, then the network is stable under the variable rate MGMW algorithm for all arrival rate vectors $\vec{\lambda}$ satisfying $\vec{\lambda} \in \sigma_M^* \Lambda$.

We consider the fluid limit model of the system as defined in the proof of Theorem 1. Consider the times when the derivative $\frac{d}{dt} q_l(t)$ exists for all $l \in \mathcal{E}$. Let $L_0(t)$ denote the set of links with the highest weight at time $t$. The weight of each link in $L_0(t)$, given by eq. (5) is denoted by $w_m(t)$. We show in Lemma 3 that $w_m(t)$ is differentiable almost everywhere. The derivative of the weights of links in $L_0(t)$ is then given by

$$\hat{w}_m(t) = \begin{cases} \frac{d}{dt} q_j(t) c_j & m \text{ is a point-to-point link } j, \\ \frac{d}{dt} \left( q_k(t) c_{kl}^*(t) + q_l(t) c_{lk}^*(t) \right) & m \text{ is a} \\ & \text{multiuser link } (k, l). \end{cases}$$

Let $L(t)$ denote the set of links from $L_0(t)$, which have the maximum derivative of the weights,

$$L(t) = \{m \in \mathcal{L} \mid \hat{w}_m(t) = \max_{i \in L_0(t)} \hat{w}_i(t)\}.$$

Then, there exists a small $\delta$ such that in the interval $(t, t+\delta)$, links in $L(t)$ will have the highest weights in $(t, t+\delta)$, *i.e.,*

$$\min_{m \in L(t)} \hat{w}_m(\tau) > \max_{m \in \mathcal{L} \setminus L(t)} \hat{w}_m(\tau). \tag{33}$$

Let $\vec{c}(\tau) \in \mathcal{C}_{L(t)}$ be the rate assignments under which the links in $L(t)$ achieve the maximum weight at times $\tau \in (t, t + \delta)$. Since variable rate MGMW picks links from $L(t)$ first, the rate allocation vector selected by variable rate MGMW, when projected on $L(t)$ will be an element of $\mathsf{R}_{L(t)}^{c(\tau)}$. We show in Lemma 4 that the service rate vector $\vec{\pi}(t)$, under variable rate MGMW, when projected on $L(t)$ is a convex combination of the elements of $\mathsf{R}_{L(t)}^{c(t)}$. Using Lemma 3, along with the fact that $\vec{\lambda}$ lies strictly within $\sigma_M^* \Lambda$., we show in Lemma 5 that there exists $m \in L(t)$ such that

$$\hat{w}_m(t) \leq -\epsilon^*, \tag{34}$$

where $\epsilon^* > 0$. Since all links in $L(t)$ have the same derivative, (34) implies that $\hat{w}_m \leq -\epsilon^*$, $\forall m \in L(t)$. Now, we can consider the following Lyapunov function $V(t) := \max_{m \in \mathcal{L}} w_m$. For $V(t) > 0$, we have that

$$\frac{d^+}{dt^+} V(t) \leq \max_{m \in \mathcal{L}} \hat{w}_m \leq -\epsilon_*. \tag{35}$$

where $\frac{d^+}{dt^+} V(t) = \lim_{\delta \downarrow 0} \frac{V(t+\delta) - V(t)}{\delta}$ is the right hand derivative of $V(t)$. This implies that the largest weight must decrease in the time interval $(t, t + \delta)$. Since the above inequality holds almost everywhere in $t$, the negative drift of the Lyapunov function implies that the fluid limit model of the system is stable and hence by the result in [11], the original system is also stable. Thus the efficiency of variable rate MGMW is atleast as large as the multiuser local pooling factor. ∎

**Lemma 3.** *For all $m \in L(t)$, the link weight $w_m(t)$ is differentiable almost everywhere in $t \geq 0$.*

*Proof:* Suppose $m$ is a point to point link $j$. The weight $w_m(t) = q_j(t) c_j$ is absolutely continuous since $c_j$ is a scalar constant and $q_j(t)$ is absolutely continuous. When $m$ is a BC link $(k, l)$, we use the Implicit Function Theorem [18] to show that $w_m(t)$ is absolutely continuous. Since $w_m(t)$ maximizes the inner product $\langle (c_{kl}, c_{lk}), (q_k(t), q_l(t)) \rangle$, the optimal rate pair can be expressed as $\eta(t) (q_k(t), q_l(t))$, where $\eta(t)$ is the proportionality factor. Using the Implicit Function Theorem, we can express $\eta$ as a continuously differentiable function of $q_k$ and $q_l$. Since $q_k(t)$ and $q_l(t)$ are differentiable almost everywhere, and by writing $w_m(t)$ as $q_k^2(t)\eta(t) + q_l^2(t)\eta(t)$, we obtain that $w_m(t)$ is also differentiable almost everywhere. ∎

**Lemma 4.** *The service rate vector $\vec{\pi}(t)$ projected on $L(t)$ is a convex combination of the elements of $\mathsf{R}_{L(t)}^{c(t)}$.*

*Proof:* We know from Eq. (33) that $\min_{m \in L(t)} \hat{w}_m(\tau) > \max_{m \in \mathcal{L}\setminus L(t)} \hat{w}(\tau)$ in $(t, t + \delta)$. This implies there exists an $n_1$ such that for all $n > n_1$,

$$\min_{i \in L(t)} w_i^n(\tau) > \max_{p \in \setminus L(t)} w_p^n(\tau).$$

Let $\vec{r}^n(\tau)$ be the rate allocation vector chosen by variable rate MGMW in time slot $\tau \in (t, t + \delta)$. Then for all $n > n_1$, $\vec{r}^n(\tau)$ projected on $L(t)$ belongs to the set $\mathsf{R}_{L(t)}^{c(\tau)}$, where $c(\tau) \in \mathcal{C}_{L(t)}$ is the transmission rate vector associated with the links in $L(t)$ at time $\tau$. Consider the service rate vector $\vec{\pi}^n$ defined in terms of the cumulative service process as:

$$\pi_l^n = \frac{S_l^n(n(t+\delta)) - S_l^n(nt)}{n\delta} \quad \text{for link } l. \tag{36}$$

We now show that $\vec{\pi}^n$ can be expressed as a convex combination of the rate allocation vectors in $\mathsf{R}_{L(t)}^{c(t)}$. We first consider an edge $l$, such that $(k, l)$ is a multiuser link in $L(t)$. From Eq. (36), $\pi_l^n$ can be expressed in terms of the rate allocation vector in each time slot in the interval $(t, t + \delta)$, so that

$$\pi_l^n = \frac{\int_{nt}^{nt+\delta} r_l^n(\tau) d\tau}{n\delta} \tag{37}$$

Considering only those time slots $\tau_i$ when $r_l(\tau)$ serves link $(k, l)$, and denoting $\alpha_1$ as the fraction of the time $l$ is served as a point-to-point link, we have

$$\pi_l^n = \frac{\sum_{i=0}^{k_1} \int_{nt+\tau_i}^{nt+\tau_{i+1}} c_{lk}^*(\tau_i) d\tau}{n\delta} + \alpha_1 c_l \tag{38}$$

$$= \frac{\sum_{i=0}^{k_1} \int_{nt+\tau_i}^{nt+\tau_{i+1}} [c_{lk}^*(\tau_i) - c_{lk}^*(t) + c_{lk}^*(t)] d\tau}{n\delta} + \alpha_1 c_l. \tag{39}$$

$$= \frac{\sum_{i=0}^{k_1} \int_{nt+\tau_i}^{nt+\tau_{i+1}} [c_{lk}^*(\tau_i) - c_{lk}^*(t)] d\tau}{n\delta} + \alpha_1 c_l + \alpha_2 c_{lk}^*(t), \tag{40}$$

where $\alpha_2 = \frac{\sum_{i=0}^{k_1} \tau_i}{n\delta}$. Since $c_{lk}^*(\tau)$ is continuous in $\tau \geq 0$, there exists an $\epsilon$ such that $|c_{lk}^*(\tau_i) - c_{lk}^*(t)| < \epsilon$. Then, for all $n > n_1$, we can bound the first term of Eq. (40) as

$$\frac{\sum_{i=0}^{k_1} \int_{nt+\tau_i}^{nt+\tau_{i+1}} [c_{lk}^*(\tau) - c_{lk}^*(t)] d\tau}{n\delta} \leq \frac{\epsilon k \max_i \tau_i}{n\delta} \leq \epsilon. \tag{41}$$

Since $c_{lk}^*(t)$ is continuous in $t$, $\forall n > n_1$, as $\delta \to 0$, we have $\epsilon \to 0$. From Eq. (40), we note that by making $\epsilon$ arbitrarily small, $\pi_l^n$ can be expressed in terms of the transmission rates in $\vec{c}(t)$ alone. Also, since the transmission rate of any point-to-point link $k$ is fixed for any $\vec{c}(\tau)$, $\pi_k^n$ can be expressed in terms of $\vec{c}(t)$. Hence, for any edge $l$ in $L(t)$, we can write $\pi_l^n$ for $n > n_1$ as,

$$\pi_l^n = \epsilon + \frac{\int_{nt}^{nt+\delta} r_l^n(\tau) d\tau}{n\delta} \tag{42}$$

where $r_l^n(\tau) \in \mathsf{R}_{L(t)}^{c(t)}$. The second term in Eq. (42) is a convex combination of the rate allocation vectors in $\mathsf{R}_{L(t)}^{c(t)}$. As $\pi_l^n \to \pi_l(t)$, $\epsilon \to 0$, and hence $\vec{\pi}(t)$ is a convex combination of the rate vectors in $\mathsf{R}_{L(t)}^{c(t)}$. ∎

**Lemma 5.** *Given that $\vec{\lambda} \in \sigma_M^* \Lambda$, there exists a link $m \in L(t)$ such that $\hat{w}_m(t) \leq -\epsilon^*$, for some $\epsilon^* > 0$.*

*Proof:* If $\vec{\lambda} \in \sigma_M^* \Lambda$, $\vec{\lambda}$ projected on the subset $L(t)$ is of the form $\sigma_M^* \mu$, where $\mu$ is a convex combination of the rate allocation vectors in $\mathsf{R}_{L(t)}$. Then, using Lemma 4 and the fact that $L(t)$ satisfies $\sigma_M^*$ local pooling, we know that there exists a link $m \in L(t)$ satisfying Eq. (44), given by:

$$\lambda_j < \pi_j(t) \qquad \text{m is point-to-point link } j, \text{ or} \tag{43}$$

$$\lambda_k < \pi_k \text{ and } \lambda_l < \pi_l(t) \qquad \text{m is a multiuser link } (k, l). \tag{44}$$

Eq. (44) implies that there exists a link $m \in L(t)$ such that

$$\hat{w}_m = \frac{dq_j}{dt} c_j = \left( \vec{\lambda}_j(t) - \vec{\pi}_j(t) \right) c_j < 0, \tag{45}$$

for point-to-point link $j$, or

$$\hat{w}_m = \frac{d}{dt} (q_k(t) c_{kl}^*(t) + q_l(t) c_{lk}^*(t)) < 0, \tag{46}$$

for multiuser link $(k, l)$. Eq. (45) follows if Eq. (43) is satisfied. We then provide the following argument to show that Eq. (44) implies Eq. (46) for a multiuser link $(k, l)$: Suppose (46) is not satisfied and $\hat{w}_m \geq 0$. Since Eq. (44) implies that $\frac{dq_k}{dt} < 0$ and $\frac{dq_l}{dt} < 0$, there exists a $\delta_1$, where $\delta_1 > 0$, such that

$$q_k(t+\delta_1) < q_j(t), \quad q_l(t+\delta_1) < q_i(t), \text{ and} \tag{47}$$

$$w_m(t+\delta_1) \geq w_m(t). \tag{48}$$

Let $(c_{kl}^*(t+\delta_1), c_{lk}^*(t+\delta_1)) = \text{argmax}_{c_{kl}, c_{lk} \in \mathcal{R}_{kl}} q_k(t+\delta_1)c_{kl} + q_l(t+\delta_1)c_{lk}$. Then, Eq. (47) implies that

$$q_k(t)c_{kl}^*(t+\delta_1) + q_i(t)c_{lk}^*(t+\delta_1) > w_m(t+\delta_1).$$

However, from our definition of weight $q_k(t)c_{kl}^*(t+\delta_1) + q_i(t)c_{lk}^*(t+\delta_1) \leq w_m(t)$, and therefore $w_m(t) > w_m(t+\delta_1)$, which contradicts Eq. (48). Hence, (46) must be satisfied. Then, by defining $\epsilon^* > 0$ similar manner as done in Eq. (18), we conclude that $\hat{w}_m(t) \leq -\epsilon^*$. ∎

## Appendix E

### Proof of Theorem 5

Suppose, for a candidate MW subset $L_{MW}$ with associated transmission rate vector $\vec{c}^o \in \mathcal{C}_{L_{MW}}$, there exists a $\hat{\sigma}_M > 0$, and a pair of vectors $\vec{\nu}$, $\vec{\mu}$ satisfying $\hat{\sigma}_M \vec{\mu} \geq \vec{\nu}$,; where $\vec{\nu}$ is a convex combination of rate vectors in $\mathsf{R}_{L_{MW}}^{\vec{c}^o}$, and $\vec{\mu}$ is a convex combination of rate vectors in $\mathsf{R}_{L_{MW}}$. Then, in a manner similar to the proof of Theorem 2, we can construct a traffic pattern with arrival rate $\vec{\lambda} = \vec{\nu} + \epsilon \vec{k}$, where $\epsilon > 0$ is arbitrary and $\vec{k} \succeq 0$ is fixed, such that the system is unstable under the variable rate MGMW policy. The arrival rate $\vec{\lambda}$ is arbitrarily close to, but outside the boundary of the region $\hat{\sigma}_M \Gamma$. Let $\hat{\vec{\nu}} = \sum_{i=0}^{J-1} \alpha_i \vec{r}_i$, and $\sum_{i=1}^{J} \alpha_i = 1$, where $\vec{r}_i \in \mathsf{R}_{L_{MW}}^{\vec{c}^o}$. We now specify the statistics of the arrival pattern in each time slot: Let the initial queue lengths $\vec{Q}(0) = 0$. At each time slot, $t \geq 1$, we pick a vector $\vec{r}_i \in \mathsf{R}_{L_{MW}}^{\vec{c}^o}$ with probability $\alpha_i$. Conditioned on $\vec{r}_i$, one of two events may occur.

**1.** With probability $1 - \epsilon$,

(a) $c_j$ packets arrive into the queue of point-to-point link $j$, $\forall j \in L_{MW}$ such that $r_i(j) = c_j$.

(b) $c^o(k)$ and $c^o(l)$ packets arrive into the queues of every multiuser link $(k, l) \in L_{MW}$ for which $r_i(k) > 0$ and $r_i(l) > 0$.

**2.** With probability $\epsilon$,

(a) $\hat{c}_j + c_j$ packets arrive into the queue of every point-to-point link $j \in L_{MW}$ for which $r_i(j) = c_j$.

(b) $\hat{c}_k + \vec{c}^o(k)$ and $\hat{c}_l + \vec{c}^o(l)$ packets arrive into the queues of every multiuser link $(k, l) \in L_{MW}$ for which $r_i(k) > 0$ and $r_i(l) > 0$.

(c) $\hat{c}_j$ packets arrive into the queues of all edges $j \in \mathcal{E}_{L_{MW}}$ such that $r_i(j) = 0$.

$\hat{c}_j$, $\hat{c}_{kl}$, and $\hat{c}_{lk}$ are fixed positive quantities defined for every point-to-point link $j \in L_{MW}$, and $(k, l) \in L_{MW}$. They satisfy the following conditions:

$$\hat{c}_j c_j = \hat{c}_k c^o(k) + \hat{c}_l c^o(l), \ \forall j, (k, l), \in L_{MW},$$

$$\hat{c}_l\, c^o(k) = c^o(l)\, \hat{c}_k, \ \forall (k, l) \in L_{MW}. \tag{49}$$

In the following lemma we show that these quantities exist.

**Lemma 6.** *There exist fixed quantities $\hat{c}_j > 0$, corresponding to every edge $j \in \mathcal{E}_{L_{MW}}$, such that they satisfy the conditions in* (49).

*Proof:* The proof of Lemma 6 is similar to the proof of Lemma 1 and follows by equating the quantities in (49) to a constant $K$ and then computing each $c_j$, $c_{kl}$ and $c_{ik}$ in terms of $K$. ∎

We now discuss the behaviour of the queues in $L_{MW}$ in Lemmas 6 and 7. As a consequence of Lemmas 6 and 7, we will show that the queues become unstable under the described arrival traffic.

**Lemma 7.** *Suppose the queues in $L_{MW}$ satisfy the condition in* (49) *as stated below:*

$$Q_j c_j = Q_k c^o(k) + Q_l c^o(l) \ \forall \ j, (k, l) \in L_{MW},$$

$$Q_k c^o(k) = c^o(l) \, Q_l, \qquad \forall \, (k, l) \in L_{MW}. \tag{50}$$

*Then, if the queue length $Q_j$, for every $j \in \mathcal{E}_{L_{MW}}$, is increased to $Q_j + \hat{c}_j$, the new queue lengths will again satisfy relation* (50).

*Proof:* From relations (49) and (50), we observe that $(Q_j + \hat{c}_j)c_j = (Q_k + \hat{c}_k)c^o(k) + (Q_l + \hat{c}_l)c^o(l)$. Also, using (50) we obtain $(Q_l + \hat{c}_l) \, c^o(k) = c^o(l) \, (Q_k + \hat{c}_k)$. ∎

**Lemma 8.** *Suppose at the end of time slot $t - 1$, the queues in $L_{MW}$ satisfy the relation* (50). *Then, at the end of time slot $t$, with probability $1 - \epsilon$, the length of all queues in $L_{MW}$ remains unchanged, and with probability $\epsilon$, the lengths of every queue in $L_{MW}$ increases by a fixed positive quantity; i.e., if $\tilde{Q}_j = Q_j(t + 1) - A_j(t)$ denotes the queue of $j \in \mathcal{E}_{L_{MW}}$ at the end of time slot $t$, then with probability $1 - \epsilon$, $\tilde{Q}_j(t) = \tilde{Q}_j(t - 1) \ \forall j \in \mathcal{E}_{L_{MW}}$, and with probability $\epsilon$, $\tilde{Q}_j(t) = \tilde{Q}_j(t) + \hat{c}_j, \ \forall j \in \mathcal{E}_{L_{MW}}$.*

*Proof:* We consider the arrival statistics of packets that we defined previously. Conditioned on vector $\vec{r}_i$ being picked, packets only arrive into those links that have non zero rates in $\vec{r}_i$. We now show that the weight of links that have a non-zero rate in $\vec{r}_i$, when calculated after the arrival of packets, strictly dominates the weight of all other links in $L_{MW}$; and hence, MGMW picks the rate allocation vector $\vec{r}_i$ . Consider first the packet arrivals as described in event **1**. Queues in $L_{MW}$, by assumption satisfy relation (50) at the end of slot $t$. At the beginning of time slot $t + 1$, after packets have arrived, the fOllowing is true: For all multiuser links $(k, l) \in L_{MW}$ satisfying $r_i(k) > 0$ and $r_i(l) > 0$:

$$(c_k^o, c_l^o) = \operatorname*{argmax}_{c_{kl}, c_{lk} \in \mathcal{R}_{kl}} \, [(\tilde{Q}_k + c_k^o)c_{kl} + (\tilde{Q}_l + c_l^o)c_{lk}], \text{ and} \tag{51}$$

$$(\tilde{Q}_k + c_k^o)c_k^o + (\tilde{Q}_l + c_l^o)c_l^o > \tilde{Q}_m c_m^o + \tilde{Q}_n c_n^o, \tag{52}$$

$\forall (m, n) \in L_{MW}$ satisfying $r_i(m) = r_i(n) = 0$. Since $\tilde{Q}_k$ and $\tilde{Q}_l$ satisfy (50), $(\tilde{Q}_k + c_k^o)$ and $\tilde{Q}_l + c_l^o)$ also satisfy (50), implying that the queue length ratios remain unchanged. (51) follows from the fact that the queue length ratios remain unchanged and the pair $(c_k^o, c_l^o)$ lies on the boundary of $\mathcal{R}_{kl}$. (52) is obtained by noting that the weight of links that are not served in $r_i$, remains unchanged. In a similar manner, for point-to-point links $j \in L_{MW}$ satisfying $r_i(j) > 0$,

$$\tilde{Q}_j(t) + c_j)c_j > \max \left( \tilde{Q}_l(t)c_l, \, \tilde{Q}_m(t)c_m^o + \tilde{Q}_n(t)c_n^o, \right) \tag{53}$$

for all links $l$, $(m, n) \in L_{MW}$ satisfying $r_i(l) = r_i(m) = r_i(n) = 0$. Relations (52) and (53) together imply that MGMW schedules rate allocation vector $r_i$ over $L_{MW}$ in time slot $t$. Since the number of packets arriving into queue $j$ equals $r_i(j)$, at the end of time slot $t$, $\tilde{Q}_j(t) = \tilde{Q}_j(t-1)$, $\forall j \in \mathcal{E}_{L_{MW}}$.

Next, consider the case when packets arrive as described in event **2**. Applying *Lemma* 7, the queue lengths $\tilde{Q}_j + \hat{c}_j$, $\forall j \in \mathcal{E}_{L_{MW}}$ will satisfy (50). Since the additional packets that arrive into each $j \in \mathcal{E}_{L_{MW}}$ equals $r_i(j)$, the same argument used for packet arrivals in event **1** implies that MGMW again picks rate allocation vector $\vec{r}_i$, and hence, $\tilde{Q}_j(t) = \tilde{Q}_j(t-1) + \hat{c}_j$, at the end of time slot $t$. ∎

We can now describe the queue evolution pattern for our packet arrival traffic. The initial queue state $\tilde{Q}_j = 0, \forall j \in \mathcal{E}_{L_{MW}}$ satisfies (50). From Lemma 8, the queue lengths do not decrease, and the event that each queue length increases by a fixed positive quantity occurs infinitely often. Hence, the system is unstable under our proposed arrival traffic. The arrival rate of the traffic pattern is then evaluated as follows: $\vec{\lambda} = \sum_{i=0}^{J-1} \alpha_i (1-\epsilon) \vec{r}_i + \sum_{i=1}^{J-1} \epsilon \vec{k}$, where $k(j) = \hat{c}_j$, $\forall j \in \mathcal{E}_{L_{MW}}$. □